**Endpoint slippage analysis in the presence of impedance rise and loss of active material**

Marco-Tulio F. Rodrigues

Chemical Sciences and Engineering Division, Argonne National Laboratory, Lemont, IL, USA

**Contact:** marco@anl.gov

**Abstract**

The endpoint slippage analysis can be used to quantify the reduction and oxidation side-reactions occurring in rechargeable batteries. Application of this technique often disregards the interference of additional aging modes, such as impedance rise and loss of active material (LAM). Here, we show that these modes can themselves induce slippage of endpoints, making the direct determination of parasitic reactions more difficult. We provide equations that describe the slippages caused by LAM and impedance rise. We show that these equations can, in principle, account for the contribution of these additional modes to endpoint slippage, enabling "correction" of testing data to quantify the side-reactions of interest. However, the challenge with this approach is that it requires information about the average $Li^+$ content of disconnected active material domains, which is, in many cases, unknowable. The present work explores mathematical connections between measurable quantities (such as capacity fade and endpoint slippages) and the extent of LAM or impedance rise endured by the cell, and discuss how the tracking of endpoints can better serve battery diagnostics.

**Introduction**

Much of the aging behavior of rechargeable batteries is determined by physical and chemical processes occurring at electrode-electrolyte interphases.[1] An example of these processes is the formation of the solid electrolyte interphase (SEI), where the reduction of certain species casts a layer that prevents additional decomposition at the surface of the negative electrode (NE). The imperfect passivation provided by the SEI causes small rates of parasitic reduction to persist throughout the cell life, constituting a major source of capacity fade.[2-4] Electrolyte components can also be oxidized at the surface of the positive electrode (PE). This parasitic oxidation injects electrons into the PE that can then recombine with $Li^+$ from the electrolyte.[4-8] This dynamic can generate capacity gain but will deplete the electrolyte of lithium, being detrimental to cell operation at high rates.[9] Interfacial reactions may also occur through the reversible conversion of redox-active species present in the cell (sometimes as contaminants), creating redox shuttles that accelerate self-discharge.[10, 11] Being able to quantify the extent of side-reactions is important to identify root causes of performance loss.

The most used method to quantify the extent of these side-reactions from experimental full-cell data is the endpoint slippage analysis.[7, 12, 13] With this technique, reduction and oxidation parasitic processes are differentiated by considering how the termination (i.e., the "endpoints") of cell charge and discharge evolves when voltage profiles are represented along a cumulative capacity axis. In traditional Li-ion cells containing graphite as negative electrode, the correspondence between the movement of endpoints (i.e., "slippage") and side-reactions is straightforward. Consider that the voltage profile for the PE and NE (solid trace) shown in Figure 1 represents the initial state of a NMC532 vs graphite cell under storage, with the vertical line indicating the utilization of the electrodes at the end of charge (EOC). Reduction side-reactions would consume electrons from the NE without affecting the state of delithiation of the PE. Hence, this type of parasitic process would cause the EOC to occur at a lower $Li^+$ content at the NE, which can be represented by sliding the NE profile to the right relative to that of the PE (dashed trace). Since the cell discharge is generally determined by the point at which the NE "polarizes" (that is, when its voltage profile becomes increasingly vertical),[4,5] this move to the right causes the future end of discharge (EOD) to occur at a higher state of delithiation of the PE. If successive charge and discharge profiles of a cell experiencing parasitic reduction are represented along a cumulative capacity axis, progressive right-shift of the discharge endpoint is seen (Figure 1b). Conceptually, this follows from the $Li^+$ consumption by the side-reactions, causing less capacity to be available during the ensuing discharge. The endpoint for charge is unaffected by reduction side-reactions, as the plateau of graphite ensures that the voltage cutoff delimiting the EOC will always occur at a same PE potential.

In the case of parasitic oxidation, the extra $Li^+$ acquired from the side-reaction are added to the PE's inventory. Consequently, additional charging is needed for the PE to reach its EOC potential, elongating the half-cycle and prompting the charge endpoint to drift rightwards (Figure 1c). Finally, for a redox shuttle, the reacting species are reversibly oxidized and reduced at the PE and NE, respectively, promoting equal shifts of both endpoints (Figure 1d). Real-world batteries will present some combination of these three prototypical processes, each producing endpoint slippages in proportion to their individual extent. We note that the direct assignment of these slippages to types of side-reactions can be affected by the identity of the electrodes and choice of voltage window.[7]

Rechargeable batteries are typically subject to more aging modes than just these three general processes.[14-16] During the cell lifetime, mechanisms such as particle fracturing, buildup of electrolyte degradation products, and electrolyte depletion can lead to inactivation of portions of the active material due to disconnection from the electronic and/or ionic pathways within the electrode.[17-20] As we discuss below, this *loss of active material* (LAM) can also promote endpoint slippage, adding a confounding factor that may prevent the attribution and quantification of reduction and oxidation side-reactions.

The interference of LAM with the endpoint slippage analysis can be quite complex, as it depends on *when* particle disconnection happens during cell aging. Dubarry et al. classified LAM into four extremes based on the target electrode and the average state of charge (SOC) of isolated domains.[14] LAM happening when the cell is charged could involve loss of delithiated PE (LAMdePE) or lithiated NE (LAMliNE). Conversely, LAM initiated at the discharged state would result in the disconnection of lithiated PE (LAMliPE) or delithiated NE (LAMdeNE) domains. While confining LAM to occur at extreme SOCs may be a simplification, this framework remains very useful for evaluating the consequences of cell aging. A key point is that the lithium content of the electrode at the time of LAM determines how much $Li^+$ will be trapped within the particles at the point of isolation. This trapped $Li^+$ becomes unavailable for reversible charge storage, meaning that LAM could directly result in capacity loss.

Figure 2 shows how the different modalities of LAM affect the charge and discharge endpoints. While LAM will generally lead to a "shrinkage" of the voltage profile of the affected electrode, the pivot point of this change will depend on the state of charge (SOC) of the cell.[14, 21] Consider LAMliNE in Figure 2a. Disconnection of NE domains when the cell is charged will result not only in loss of active domains that can host $Li^+$, but also to loss of the bound lithium ions themselves. The subsequent discharge half-cycle will terminate at a higher PE delithiation state, as the NE is unable to return the same amount of $Li^+$ it had originally received. Consequently, the cell will exhibit a rightward slippage of the discharge endpoint (Figure 2b), much like it would be observed for parasitic reduction (Figure 1b). Note that LAMliNE will not affect the Li$^+$ content of PE and NE at which the cell reaches the EOC; although it decreases the number of active domains available at the NE, the relative utilization of these domains upon charging the cell remains the same. This invariance of EOC conditions causes the charge endpoint to remain static (Figure 2b).

In contrast, LAMdeNE affects the state of the electrodes at the EOC but not at the EOD (Figure 2c), resulting in invariance of the discharge endpoint (Figure 2d). Shifts to the charge endpoint can be subtle and will depend on the extent of LAM and on features of the voltage profiles of the positive and negative electrodes. Charging a cell that has suffered LAMdeNE will lead to an increased utilization of the remaining active NE domains. That is, in a graphite electrode, the EOC will occur further along its plateau. The constancy of the NE potential along this plateau effectively "pins" the EOC at the same PE potential, fixing the charge endpoint of the cell. At advanced stages of LAMdeNE, however, over-utilization of the NE during charging may eventually lead to the EOC occurring beyond the graphite plateau, limiting the delithiation achieved by the PE at the end of charge. When this transition beyond the plateau occurs, the charge endpoint will start receding leftwards (Figure 2b), countering the trends exhibited by parasitic oxidation (Figure 1c). (For cases where the NE does not exhibit a plateau, such as silicon, changes to the charge endpoint would not display this "incubation period".)

A similar thought process can be applied to understand the effects of loss of active PE material. For LAMliPE, disconnection of portions of the PE will also result in loss of the Li$^+$ within. As LAM progresses, the remaining active PE domains will provide a progressively lower amount of Li$^+$ to the NE before reaching the electrode potentials corresponding to the EOC (Figure 2e). This shortens the charge half-cycle and causes a leftward slippage of the charge endpoint

(Figure 2f). In the case of LAMdePE, observation of slippage is also subject to an incubation period, being negligible until sufficient LAM has occurred to cause the EOD to become "limited" by polarization of the PE (Figure 2g). When that happens, the discharge endpoint will shift to the right (Figure 2h). In conclusion, LAM can result in shifts of the endpoints that either amplify or counteract the effects caused by oxidation and reduction side-reactions.

Another common consequence of cell aging is impedance rise. Increases in cell resistance can often be dominated by interfacial changes at the PE.[8, 20] Later in life, the same mechanisms responsible for LAM (i.e., electrolyte dry-out, and pore clogging due to excessive SEI growth) can also contribute to higher impedances.[18, 22] At first approximation, moderate impedance rise acts by increasing the polarization of the electrodes. Figure 2i shows an example in which the impedance of the PE has increased, imposing an overpotential of 50 mV. Much like shown in Figure 2e for LAMliPE, this increased polarization will cause the cell to reach the EOC earlier, leading to a leftward slippage of the charge endpoint (Figure 2j). Note that, since the cell voltage is the point-by-point difference between the PE and NE potentials, the outcome is the same regardless of which electrode experiences the added polarization. Effects on the discharge endpoint are more muted, as changes in polarization are neutralized by the steep voltage profile of the NE that determine the occurrence of the EOD.

The examples of Figure 2 highlight the complexities of how these additional aging modes can affect the position of the endpoints. Modes affecting the charge endpoint (LAMdeNE, LAMliPE and impedance rise) are especially critical. This endpoint is used to identify rates of parasitic oxidation (Figure 1c), and since these side-reactions are typically less prevalent than parasitic reduction,[4, 12, 23, 24] any external interference from these other modes could have a large relative impact on the interpretation of experimental trends.

The extent of impedance rise and LAM experienced by a cell are accessible from experimental data. The former can be directly measured; the latter can be inferred from differential voltage analysis (DVA) or similar methods.[14, 15] We wondered whether this information could be used to quantify the contributions from these modes to endpoint slippage, which could then be subtracted from the experimental data to enable accurate determination of the rates of parasitic oxidation and reduction. This interest arose from our investigations of rapid calendar aging in silicon electrodes.[25] The long-term instability of Si-containing systems appears to originate from the improper protection offered by the SEI, likely leading to premature cell failure due to electrolyte starvation.[25, 26] It is thus desirable to obtain quantitative insights on how well novel electrolyte classes and compositions can passivate silicon surfaces over time. LAM has been reported for Si particles during extended storage at high SOCs (refs. [26, 27]) and must then be considered when analyzing extended aging data. Additionally, we have previously discussed how features of the voltage profile of silicon can complicate the application of endpoint slippage analysis.[7] Differently from graphite, Si-rich NEs may experience slippage of *both* charge and discharge endpoints following parasitic reduction or oxidation.[7] Consequently, knowledge of both endpoints may be required to quantify the rate of SEI growth.

Here, we show that the endpoint slippages produced by impedance rise and by loss of active material can be calculated. Contributions from these modes are then used to determine the true extent of parasitic oxidation and reduction experienced by hypothetical cells, even when they are subject to complex aging profiles. Nevertheless, we also argue that a key piece of information required for such calculations – the average $Li^+$ content of the electrode when LAM happens – is often unknown and perhaps even unknowable in many cases. Despite this challenge, this work explores mathematical relationships between causes and effects of aging that can be useful to future research.

*Experimental and Simulation*

Voltage profiles employed in simulations were obtained from tests vs. Li metal at very slow rates (< C/100); see ref. [7] for experimental details. For the simulation of aging, Python was used to apply the relevant transformations on the voltage profiles of the PE and NE. Full-cell profiles were then obtained from the point-by-point difference between electrode potentials within the assumed voltage cutoff (3 – 4.2 V). All simulations were performed in the frame of reference of the SOC scale of the unaged PE. A capacity ratio of 0.9 between the NE and PE was assumed for the simulated cells. This ratio was calculated based on the arbitrary voltage range of the half-cell data used in the simulations (0.01 – 1.0 V for the NE, 3.0 – 4.5 V for the PE). Here, overcharge was averted due to the lower utilization of the PE capacity in the full-cell. An initial offset between the PE and NE profiles of 0.05 was assumed to account for capacity losses during formation cycles.

Simulation of parasitic reduction and oxidation is discussed in detail in ref. [7]. For LAM, loss of active domains was assumed to occur either at the EOC or at the EOD, but not at any other point in between, following the conceptualization by Dubarry et al. (ref. [14]). LAM was simulated in such a way that the Li$^+$ content of the electrodes at the termination of the relevant half-cycle remained constant. For example, when simulating LAMliPE, the PE profile is slightly shifted to the right to preserve the EOD, ensuring that no shift occurs at the discharge endpoint; this can be observed in Figure 2e, for example. This approach ensures that the simulated LAM leads to loss of portions of the active material without disrupting the functioning of the remaining domains.

Impedance rise was always assumed to occur at the PE prior to the onset of charging; that is, the impedance rise in cycle $k$ affects the electrode polarization in cycle $k$.

*Results and Discussion*

*Quantification of aging from endpoint slippages.* The slippages $C_{slip}$ and $D_{slip}$ produced by various aging modes between two cycles of consideration are summarized in Table 1. Equations for parasitic reduction and oxidation are discussed in ref. [7], while derivations for the others are provided here in Sections S2-S6 of the Supplementary Material. Importantly, all equations depend on the parameters $\lambda$ and $\omega$, defined as

$$\lambda \equiv \frac{\left.\frac{dU_{PE,d}}{dq}\right|_{EOD}}{\left(\left.\frac{dU_{PE,d}}{dq}\right|_{EOD} - \left.\frac{dU_{NE,d}}{dq}\right|_{EOD}\right)} \tag{1}$$

$$\omega \equiv \frac{\left.\frac{dU_{NE,c}}{dq}\right|_{EOC}}{\left(\left.\frac{dU_{PE,c}}{dq}\right|_{EOC} - \left.\frac{dU_{NE,c}}{dq}\right|_{EOC}\right)} \tag{2}$$

The derivatives are the slopes of the voltage profiles of the PE and NE at the ends of discharge (d) and charge (c). These quantities define by how much the EOC and EOD are limited by polarization of the NE ($\omega$) and PE ($\lambda$).[5-7] From these definitions, it follows that $0 \leq \lambda \leq 1$ and $-1 \leq \omega \leq 0$; see the Supplementary Material for sign conventions. We have discussed previously why these parameters carry the necessary information to quantify consequences of aging, such as capacity fade and endpoint slippages.[5-7] Since these measurable quantities are determined by the ending of a half-cycle, they can be fully described using properties of the electrode profiles at those specific regions; the rest of the voltage curves plays a minor role in their determination. Inspection of eqs.

1 and 2 indicate that, for systems displaying strong polarization of the PE at the EOC and of the NE at the EOD (such as many graphite-based cells cycled between 0 and 1 SOC), $\lambda \approx \omega \approx 0$, simplifying the equations in Table 1.

An important property of those equations is that they are *additive*. For example, the discharge endpoint between arbitrary cycles $k-1$ and $k$ caused by a combination of parasitic reduction and LAMliNE is given by

$$D_{slip,k} = 2q_{red,k}(1-\lambda) + N_0 L_{liNE,k}(x_{EOC,k-1} - x_{EOD,k-1})(1-\lambda) \qquad (3)$$

If information about aging is available (e.g., the amount of LAM $L_{liNE,k}$ experienced by the cell), one can, in principle, write similar equations for both $C_{slip,k}$ and $D_{slip,k}$. This results in a system of two equations and two unknowns ($q_{red,k}$ and $q_{ox,k}$), which can then be solved to determine the incremental parasitic capacities experienced by the cell. We give a detailed example of these steps below; for a simpler application of these ideas when only $q_{red}$ and $q_{ox}$ are present, see ref. [7].

Table 1 also lists equations for the discharge capacity loss $Q_{loss,k}$ between cycles $k-1$ and $k$ inflicted by each aging mode. Since the accessible cell discharge capacity is given by the number of electrons exchanged in between the charge and the discharge endpoints at a given cycle, changes in discharge capacity can be expressed from the slippage of endpoints as

$$Q_{loss,k} = D_{slip,k} - C_{slip,k} \qquad (4)$$

Notably, equations for $Q_{loss,k}$ are also additive, enabling the quantification of the consequences of a complex mixture of aging modes. We have verified this additivity of $D_{slip,k}$, $C_{slip,k}$ and $Q_{loss,k}$ by observing that attempts to determine expressions for multiple aging modes (following the same logic used in Sections S2-S6 of the Supplementary Material and on refs. [6,7]) resulted in a simple sum of the equations obtained for the cases with a single aging mode. This verification has been omitted for brevity but is demonstrated graphically below.

The equations in Table 1 may require knowledge of the slopes of voltage profiles at the EOC and EOD, of the extent of LAM, and of the $Li^+$ content $x$ at the NE or the state of delithiation $y$ of the PE at the end of the relevant half-cycles. Such information can be obtained from DVA or from 3-electrode tests.[14, 15, 28] In the case of impedance rise, the current $I$ is known and the increase in cell resistance $\Delta R$ can be directly measured. As we discuss below, equations in Table 1 are valid if aging does not lead to large changes in $\omega$, $\lambda$ or of the $Li^+$ content of the electrodes at the EOC/EOD within the cycles of consideration.

All the equations above are valid within the frame of reference of the initial PE SOC scale (like used in Figure 1a and Figure 2a). Conversion to a cell SOC scale (such as in Figure 2b) requires multiplication by simple conversion factors, which are discussed in section S8 of the Supplementary Material. Throughout this manuscript, aging rates and capacity losses are presented in terms of the initial PE SOC scale, but endpoint slippages are reported with respect to the initial cell SOC axis. Assuming aging rates (especially LAM) within an electrode SOC scale is more intuitive, and also more aligned with the way these quantities are quantified experimentally (e.g., LAM is typically reported as a percentage of the initial electrode capacity). Capacity loss was maintained at this same scale to enable a direct comparison with the aging rates. However, since endpoint slippage is inherently a cell-level property, it is reported here with respect to the initial cell SOC scale.

A final note about the equations in Table 1 relates to the quantities $x$ and $y$. These $Li^+$ contents do not need to be determined in absolute terms (e.g., by considering the theoretical total amount of $Li^+$ that can be removed from a PE material). Instead, it can simply be scaled based, for example, on the baseline half-cell data used for fitting the data during DVA. The quantity $N_0$ is the initial N/P ratio of the cell (that is, the capacity ratio between the NE and PE, at the beginning of life, within the voltage window for each half-cell dataset). Multiplication of $x$ terms by $N_0$ ensures that the cell SOC window defined by these quantities is the same regardless of the voltage limits assumed for voltage profiles of each electrode. A graphical example of how the $x$ and $y$ scales are arbitrary is provided in Section S9 of the Supplementary Material.

***Decoupling $q_{red}$ and $q_{ox}$ from other aging modes.*** Consider an NMC532 vs graphite cell, like the one represented in Figure 2, and that it experiences a single aging mode at the rate specified in Figure 3. The assumed aging comprises either impedance rise (right y-axis) or a type of LAM (left y-axis), all evolving quadratically with respect to cycle number. The effects of each of these cases of simulated aging on the cell are shown in Figure 4, with consequences of each mode represented along a separate column (see header for identification of the respective mode). The onset and magnitude of capacity fade will depend on whether the aging mode leads to direct Li$^+$ loss (due to disconnection of lithiated domains) or changes to the electrode limiting the end of half-cycles (Figure 2). Note that parasitic reduction or oxidation are not applied to these cells; we will show how the equations in Table 1 can use the simulated endpoint slippages to infer that these side-reactions are indeed negligible.

Taking LAMliNE as example (leftmost column in Figure 4), it can be seen that it results in continuous capacity fade (Figure 4a) and discharge endpoint slippage (Figure 4f). At each cycle $k$, one must determine $L_{liNE,k}$ (that is, the amount of LAM accrued between cycles $k-1$ and $k$), the NE Li$^+$ content at both the EOC and EOD, and the parameters $\lambda$ and $\omega$. With this information, the system of equations of $C_{slip,k}$ and $D_{slip,k}$ can be solved at each cycle for $q_{red,k}$ and $q_{ox,k}$:

$$\begin{cases} D_{slip,k} = 2q_{red,k}(1-\lambda) + 2q_{ox,k}\lambda + N_0 L_{liNE,k}(x_{EOC,k-1} - x_{EOD,k-1})(1-\lambda) \\ C_{slip,k} = -\omega(q_{red,k} + q_{red,k-1}) + (1+\omega)(q_{ox,k} + q_{ox,k-1}) \end{cases} \quad (5)$$

As mentioned above, for this NMC532 vs graphite cell, $\lambda \approx \omega \approx 0$ (Figure 4k), greatly simplifying these equations to

$$\begin{cases} D_{slip,k} \approx 2q_{red,k} + N_0 L_{liNE,k}(x_{EOC,k-1} - x_{EOD,k-1}) \\ C_{slip,k} \approx (q_{ox,k} + q_{ox,k-1}) \end{cases} \quad (6)$$

Solving the system of equations for each cycle $k$ correctly identifies that negligible $q_{red}$ and $q_{ox}$ were applied to the system (Figure 4p). That is, all the endpoint slippage observed in Figure 4f can be fully explained by the amount of LAMliNE inflicted onto the cell. Knowing that $q_{red,k} \approx q_{ox,k} \approx 0$, we can use the $Q_{loss,k}$ equation for LAMliNE to determine the amount of capacity fade, per cycle, that is caused by the loss of active material (cumulative values shown as gray markers in Figure 4a), correctly matching the decay profile from the simulation.

Similar outcomes are obtained for the other aging modes (Figure 4): capacity fade is well-described by the equations in Table 1, and negligible $q_{red}$ and $q_{ox}$ are identified. Errors for the latter start becoming meaningful at advanced stages of LAMdePE (Figure 4s), when cell discharge becomes highly PE-limited (Figure 2g). As explained by Tornheim and O'Hanlon,[4] $q_{red}$ is not determinable in these conditions. Inspection of Table 1 shows that, at this point $\lambda \to 1$, causing the coefficients of $q_{red}$ terms in the equations for parasitic reduction to approach zero. When $1 - \lambda \approx \omega \approx 0$, it becomes impossible to evaluate $q_{red}$.

Although Figure 4 exemplifies how the equations in Table 1 could be used to extract $q_{red}$ and $q_{ox}$ from observed endpoint slippages, it requires analysis of *every* cycle, which would be quite taxing to implement. Testing of rechargeable batteries typically involves occasional reference performance tests (RPTs) to quantify the effects of aging during long experiments. Determination of LAM through analysis of voltage profiles is typically performed only on data from the RPTs.[15, 28, 29] We have previously demonstrated that $q_{red}$ and $q_{ox}$ can be determined by taking $D_{slip,k}$ and $C_{slip,k}$ as the endpoint slippages accrued in between successive RPTs (or any other arbitrary pair of non-consecutive cycles) when only these two side-reactions are present.[7] Here, we show that this remains mostly true when LAM and impedance rise occur. Figure 5 shows the same cases of Figure 4, but considering only one data point every 100 cycles. In most cases, negligible rates of parasitic reduction and oxidation are correctly inferred from the endpoint slippages (Figure 5p,r,t). However, significant deviations occur for LAMdeNE and LAMdePE (here, earlier than in Figure 4s). The problem arises from the large variations in $\omega$ and $\lambda$, respectively, observed when cell limitation changes (Figure 5l,n). Despite the rapid growth of these parameters in Figure 4, the density of data points causes the incremental values per cycle to be manageable. Here, however, the cumulative changes experienced within the 99 cycles separating the RPTs violate the assumption of near-constancy of these parameters used for deriving the equations. Although possible,[30] variations of this magnitude are not commonly observed experimentally and only tend to occur later in cell life, indicating that the analysis remains viable for most cases. Note that the deviations are emphasized by the narrow y-axis scale of Figure 5p,s;

trends for capacity loss (Figure 5b,d) offer a better perspective of the moderate limitations of the equations under these conditions.

***Endpoint slippages from multiple aging modes.*** Although Figure 4 is instructive, batteries will rarely experience a single aging mode. Hence, the same cell used in the examples above was exposed to the combined impact of parasitic reduction and oxidation, LAMliPE and LAMliNE, to the cumulative extent detailed in Figure 6a. Additionally, it was also considered that impedance rise increased PE polarization by the amount indicated in Figure 6b. The combined effect of these modes produced >20% of capacity fade (Figure 6c), and large slippages of discharge and charge endpoints (dashed black lines in Figure 6d,e, respectively).

Despite the added complexity brought by the multiple aging modes, the equations in Table 1 can correctly capture trends for capacity loss and endpoint slippages (silver markers in Figure 6c,d,e). If the necessary parameters are known, the relevant equations can again be combined and solved for $q_{red}$ and $q_{ox}$. Repeating this process cycle by cycle, true rates of oxidation and reduction can once more be correctly inferred from the slippages (Figure 6f).

Individual modal contributions to the observed shifts in endpoints are represented graphically in Figure 6d and Figure 6e for discharge and charge, respectively. Figure 6d shows that parasitic reduction and LAMliNE are the major contributors to the discharge endpoint slippage, though impedance rise also have a minor impact (refer to color code in *panel a*). In the case of charge (Figure 6e), the observable behavior results from a combination of parasitic oxidation (which moves the endpoint towards more positive values), and LAMliPE and impedance rise (that pulls it to the opposite direction). Parasitic reduction also plays a minor role, since $\omega$ slightly differs from zero (Table 1 and Figure 4k). Overall, disregarding LAM and impedance rise would lead the endpoint slippage analysis to overestimate parasitic reduction and underestimate parasitic oxidation.

It is also instructive to analyze how each aging mode affects cell capacity. Figure 7 shows that the total fade is a balance between the extra electrons accrued from parasitic oxidation versus the capacity loss resulting from the remaining aging modes. Among the latter, parasitic reduction is by far the dominant mechanism, while impedance rise still contributed to ~17.3% of the total observed loss.

The usage exemplified here for the equations in Table 1 presumes an independence between the consequences of the various aging modes. That is, it considers that the capacity lost due to LAMliNE is the same whether parasitic reduction is present. This assumption, however, overlooks interactions between the modes. For example, while LAMliNE does not affect the $Li^+$ content of the NE at the EOC, parasitic reduction does, and the concurrence of these modes decreases the direct $Li^+$ trapping upon losing active material. Without accounting for this type of interaction, the endpoint slippages and capacity loss are slightly overestimated, resulting in the minor errors seen at later cycles in Figure 6c-f. A more general set of equations for the effects of LAM is presented in Section S7 of the Supplementary Material. We note that deviations observed in Figure 6 and Figure 7 are sufficiently small that the added accuracy of the alternative expressions may not justify the added complexity.

For completeness, Figure S11 exhibits the effect of the same multi-modal aging of Figure 6 on a cell containing a silicon NE. In this case, that same aging profile led to a slightly lower capacity fade (Figure S11c), partly due to the "reservoir effect" that enables access to dormant $Li^+$ within the NE as side-reactions progress.[6] Additionally, there is a significant increase in the contribution of impedance rise and parasitic oxidation to the discharge endpoint slippage (Figure S11d), and of parasitic reduction to the charge (Figure S11e). In this silicon cell, LAMliPE and LAMliNE have a smaller impact on capacity fade, while that of impedance rise becomes higher. In fact, the contributions of impedance to capacity loss amounts to ~22.9% of the total observed fade. These differences are caused by the shape of the voltage profile of silicon, where a less steep profile at low $Li^+$ content and the lack of a plateau at high results in progressive variation of electrode potentials at the EOC and EOD as the cell ages. Such changes are captured by the parameters $\lambda$ and $\omega$ (eq. 1 and eq. 2), which present larger magnitude here than in the graphite cell (Figure S10). Despite this difference, the equations in Table 1 remained sufficiently accurate to determine the true rates of parasitic reduction and oxidation (Figure S11f).

***Practical concerns.*** The examples in the previous section highlight how it is possible to determine the true rates of $q_{red}$ and $q_{ox}$ from the endpoint slippages, even under a complex aging profile. This determination works satisfactorily when all required information is available with perfect accuracy, as was the case in our simulated examples. How realistic is the expectation that this would also apply to testing data from rechargeable batteries?

    Using the equations in Table 1 may require the knowledge of:

i.    The magnitude of impedance rise;

ii.    The $Li^+$ content $x$ of the NE at the EOC and EOD;

iii.    The delithiation extent $y$ of the PE at the EOC and EOD;

iv.    The parameters $\lambda$ and $\omega$ (that is, the slopes of the voltage profiles of the PE and NE at the EOC and EOD);

v. The magnitude and modality of LAM.

Requirement *i* is the easiest to fulfill, as it is directly measurable. Notwithstanding, impedance effects could be negligible in several cases of interest. In the examples above we assumed that the PE in the hypothetical cell would experience cumulative polarization gains of 50 mV due to increases in resistance. Considering resistance and capacity values for certain commercial cells in the literature,[31, 32] this would amount to an impedance rise of ~150% if data is collected at C/3, which is a high level of aging. If we consider that many aged cells experience impedance rise of < 50% (leading to polarization gains < 15 mV),[2, 3, 31, 33-35] direct contributions of impedance rise to $Q_{loss}$ and endpoint slippages become negligible.

The other four requirements listed above necessitate the availability of deeper knowledge about the state of the cell, from using a reference electrode, or performing DVA or similar analyses. Errors in determining $x$ and $y$ should be relatively small. Determining $\lambda$ and $\omega$ in cells at advanced stages of aging could be a bit more complex, due to possible distortions in the voltage profiles. Still, deviations may not be large enough to significantly affect the outcomes of the analysis.

The biggest challenge, however, is to decisively identify the modality of LAM. In principle, this can be accomplished from inspection of changes in the signature of the evolution of dQ/dV and dV/dQ traces,[14, 30, 36-39] as they may present subtle differences between one case and the other. Nevertheless, these differences are often too subtle to enable a robust determination of the $Li^+$ content of electrode domains at the time of disconnection. Indeed, there are just a few clear examples of such identifications in the literature,[15, 29, 30, 36-40] and they often require many layers of reasoning and the correlation of data from different sources. Achieving this differentiation is in fact so difficult that more recent works have resorted to ascribing all LAM as being in the delithiated state and adding LLI to represent $Li^+$ that could be trapped within the lost particles.[16,

[41, 42] In this manner, $LAMliNE = LAMdeNE + LLI = LAMdeNE + q_{red} - q_{ox}$. The challenge with this approach is that the Li$^+$ lost from LAMliNE cannot be differentiated from that lost due to parasitic reactions, and $q_{red}$ or $q_{ox}$ cannot be determined. Alternatively, one could attempt to directly differentiate types of LAM experimentally, though that could require resource-intensive characterization. To complicate matters further, we have recently observed that disconnected domains can self-discharge during cell storage,[43] casting doubt on the reliability of experimentally distinguishing between modalities of LAM.

Despite this fundamental limitation, there are ways in which the endpoint slippage analysis can be applied with confidence. Many commercial-grade systems do not tend to experience significant LAM for much of their lives if kept under moderate environmental conditions,[2, 3, 23, 24] eliminating the problem we discuss above. More generally, application of this analysis should always be preceded by quantification of LAM (such as done in refs. [23, 24]), and its use avoided when material disconnection becomes a relevant aging mode.

Another option is to constrain the application of endpoint slippage analysis to the beginning of the cell life (as in refs. [11, 44-46]). It is reasonable to expect that LAM would not occur extensively at this point, and so testing data can be used to directly investigate the levels of $q_{red}$ and $q_{ox}$ as part of electrolyte development.

Finally, in cases where sufficient knowledge about the system exists, it may be possible to determine the type of LAM occurring in the cell. For example, in Ni-rich layered oxide PEs LAMdePE appears to be more likely than LAMliPE, due to intergranular fracturing caused by anisotropic lattice changes at high levels of delithiation.[17, 47] In these cases, the equations in Table 1 can be used to account for the effects of LAM on the measured endpoint slippages, so that $q_{red}$ or $q_{ox}$ can be determined.

*Conclusions.*

The endpoint slippage analysis can be extremely useful for developing electrolyte systems with lower overall reactivity, identifying the existence of redox shuttles and understanding trends in long-term aging data. Nevertheless, we show here that loss of active material can also produce shifts in the endpoints, which may potentially affect the usability of this technique. With that in mind, we set to explore whether these LAM contributions to slippages could be accounted for and numerically subtracted from experimental values to reveal the true rates of underlying parasitic reduction and oxidation. We found that such an operation is indeed possible and presented equations describing how the different modes of LAM will affect the endpoints.

A useful property of these equations is that they are additive, enabling the combination of individual expressions to analyze the behavior of cells under the simultaneous influence of multiple aging modes. We provide examples showing how this enables the decoupling of contributions from these various aging modes to endpoint slippages and to capacity fade, providing quantitative insights into how each mechanism is affecting measurable quantities.

The caveat is that using the equations requires knowledge of various parameters of the cell, including the $Li^+$ content and slopes of the voltage profiles of the electrodes at the end of charge and discharge, and the extent of LAM experienced by the PE and NE. These quantities are generally accessible through analysis of testing data, using methods such as differential voltage analysis. However, applying the equations provided here also necessitates knowing *when* LAM has happened; that is, whether disconnection of active material domains happened in the lithiated or delithiated state. Judging from the lack of reliable differentiation between these cases in the literature, we conclude that this level of granularity may be difficult to achieve in many systems of interest or, at the very least, could require extensive experimental work to be determined.

With these considerations in mind, we conclude that the endpoint slippage analysis may be more suited to systems and conditions where LAM and impedance rise have not occurred to a significant extent. In other words, the technique is most effective in scenarios where the equations reported in the present work are not needed. Despite the difficulty in applying these equations, they offer a detailed analytical description of consequences of various aging modes to cell performance and can be useful in future efforts in modeling the behavior of rechargeable batteries.

*Acknowledgements*

The author is grateful to the colleague Adam Tornheim for his helpful comments. This research was supported by the U.S. Department of Energy's Vehicle Technologies Office under the Silicon Consortium Project, directed by Brian Cunningham, Thomas Do, Nicolas Eidson and Carine Steinway, and managed by Anthony Burrell. The submitted manuscript has been created by UChicago Argonne, LLC, Operator of Argonne National Laboratory ("Argonne"). Argonne, a U.S. Department of Energy Office of Science laboratory, is operated under Contract No. DE-AC02-06CH11357. The U.S. Government retains for itself, and others acting on its behalf, a paid-up nonexclusive, irrevocable worldwide license in said article to reproduce, prepare derivative works, distribute copies to the public, and perform publicly and display publicly, by or on behalf of the Government.

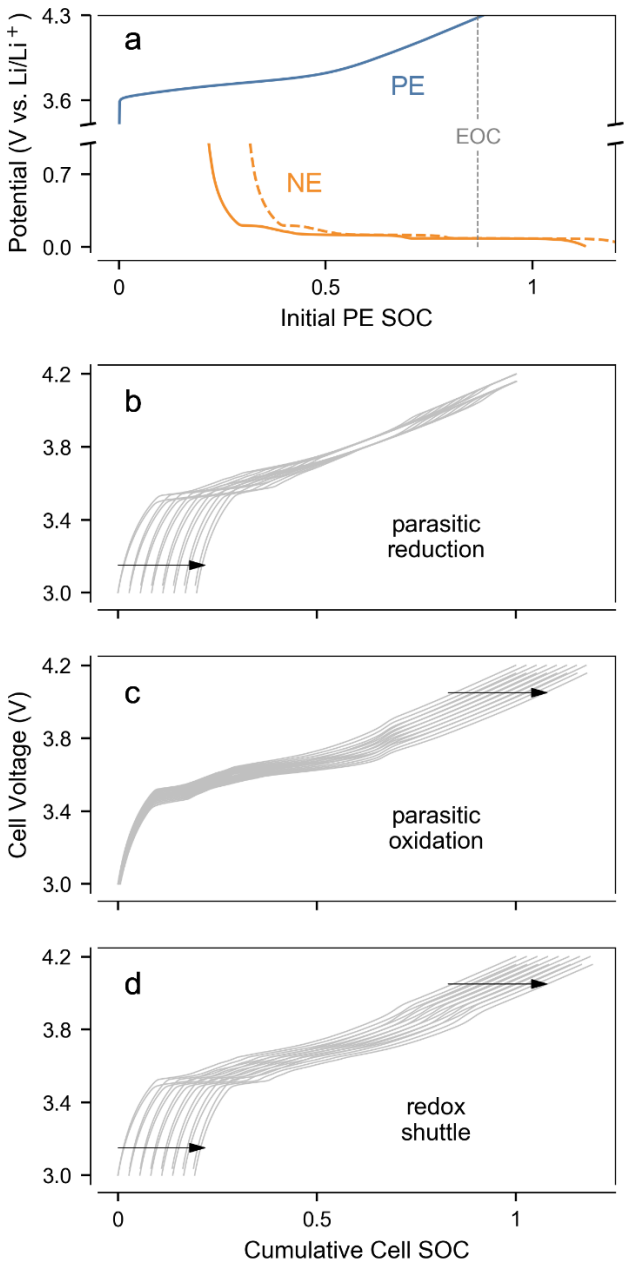

Figure 1. Overview of the endpoint slippage analysis. a) Voltage profiles for the positive and negative electrodes that define a full-cell. The dashed orange line represents the profile for the NE after aging. The end of charge point of the hypothetical full-cell is indicated by the vertical gray line. High-voltage portions of the PE profile are omitted. The other panels show examples of the endpoint slippage observed as a result of parasitic reduction (b), parasitic oxidation (c) or a redox shuttle (d).

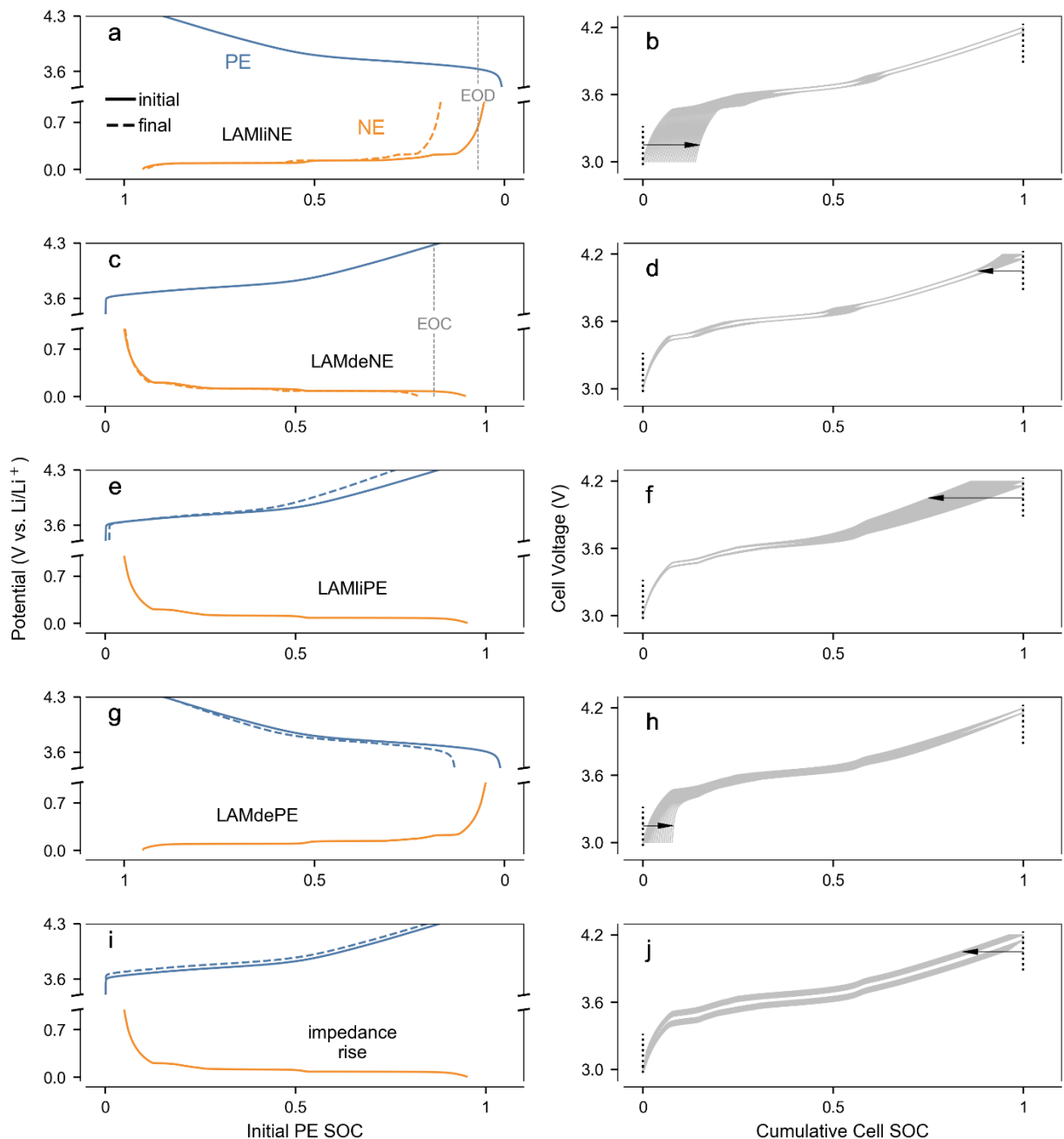

Figure 2. Simulated endpoint slippage caused by: a,b) LAMliNE; c,d) LAMdeNE; e,f) LAMliPE; g,h) LAMdePE; i,j) impedance rise. Left panels show how the various aging modes affect the voltage profiles of the electrodes, while panels to the right show the effect on full-cell data. Profiles in dashed lines in panels a,c,e,g,i indicate the final state of the electrode targeted by each aging mode. Arrows on panels b,d,f,h,j highlight the slippage of endpoints caused by aging. Vertical lines in panels a and c indicate the initial EOD and EOC, respectively. High-voltage portions of the PE profile are omitted.

Table 1. Quantification of the aging caused by the indicated aging modes. The equations determine the endpoint slippages and capacity loss in between arbitrary cycles $k-1$ and $k$ of a cell. The cumulative effect of aging is obtained by summation of these quantities over all cycles of interest. See text for definition of terms.

| Aging mode | Discharge slippage ($D_{slip,k}$) | Charge slippage ($C_{slip,k}$) | Capacity loss ($Q_{loss,k}$) |
|---|---|---|---|
| Parasitic reduction | $2q_{red,k}(1-\lambda)$ | $-\omega(q_{red,k}+q_{red,k-1})$ | $q_{red,k}(2-2\lambda+\omega)+\omega q_{red,k-1}$ |
| Parasitic oxidation | $2q_{ox,k}\lambda$ | $(1+\omega)(q_{ox,k}+q_{ox,k-1})$ | $-q_{ox,k}(1-2\lambda+\omega)-q_{ox,k-1}(1+\omega)$ |
| LAMliNE | $N_0 L_{liNE,k}(x_{EOC,k-1}-x_{EOD,k-1})(1-\lambda)$ | 0 | $N_0 L_{liNE,k}(x_{EOC,k-1}-x_{EOD,k-1})(1-\lambda)$ |
| LAMdeNE | 0 | $N_0 L_{deNE,k}\omega(x_{EOC,k-1}-x_{EOD,k-1})$ | $-N_0 L_{deNE,k}\omega(x_{EOC,k-1}-x_{EOD,k-1})$ |
| LAMliPE | 0 | $-L_{liPE,k}(y_{EOC,k-1}-y_{EOD,k-1})(1+\omega)$ | $L_{liPE,k}(y_{EOC,k-1}-y_{EOD,k-1})(1+\omega)$ |
| LAMdePE | $L_{dePE,k}\lambda(y_{EOC,k-1}-y_{EOD,k-1})$ | 0 | $L_{dePE,k}\lambda(y_{EOC,k-1}-y_{EOD,k-1})$ |
| Impedance rise | $\dfrac{-I\Delta R_k}{\left.\dfrac{dU_{cell,d}}{dq}\right|_{EOD}}$ | $\dfrac{-I\Delta R_k}{\left.\dfrac{dU_{cell,c}}{dq}\right|_{EOC}}$ | $I\Delta R_k\left(\dfrac{1}{\left.\dfrac{dU_{cell,c}}{dq}\right|_{EOC}}-\dfrac{1}{\left.\dfrac{dU_{cell,d}}{dq}\right|_{EOD}}\right)$ |

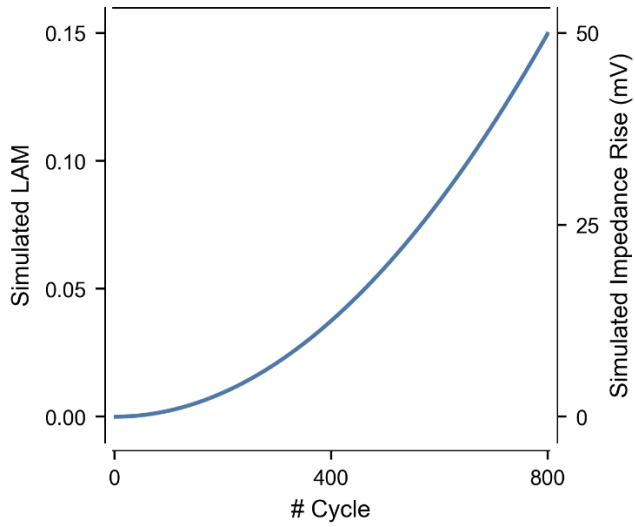

Figure 3. Aging imposed on a hypothetical cell, showing the cumulative loss of active material (left axis) or impedance rise (right) used in simulations. LAM is in units of the initial PE SOC scale.

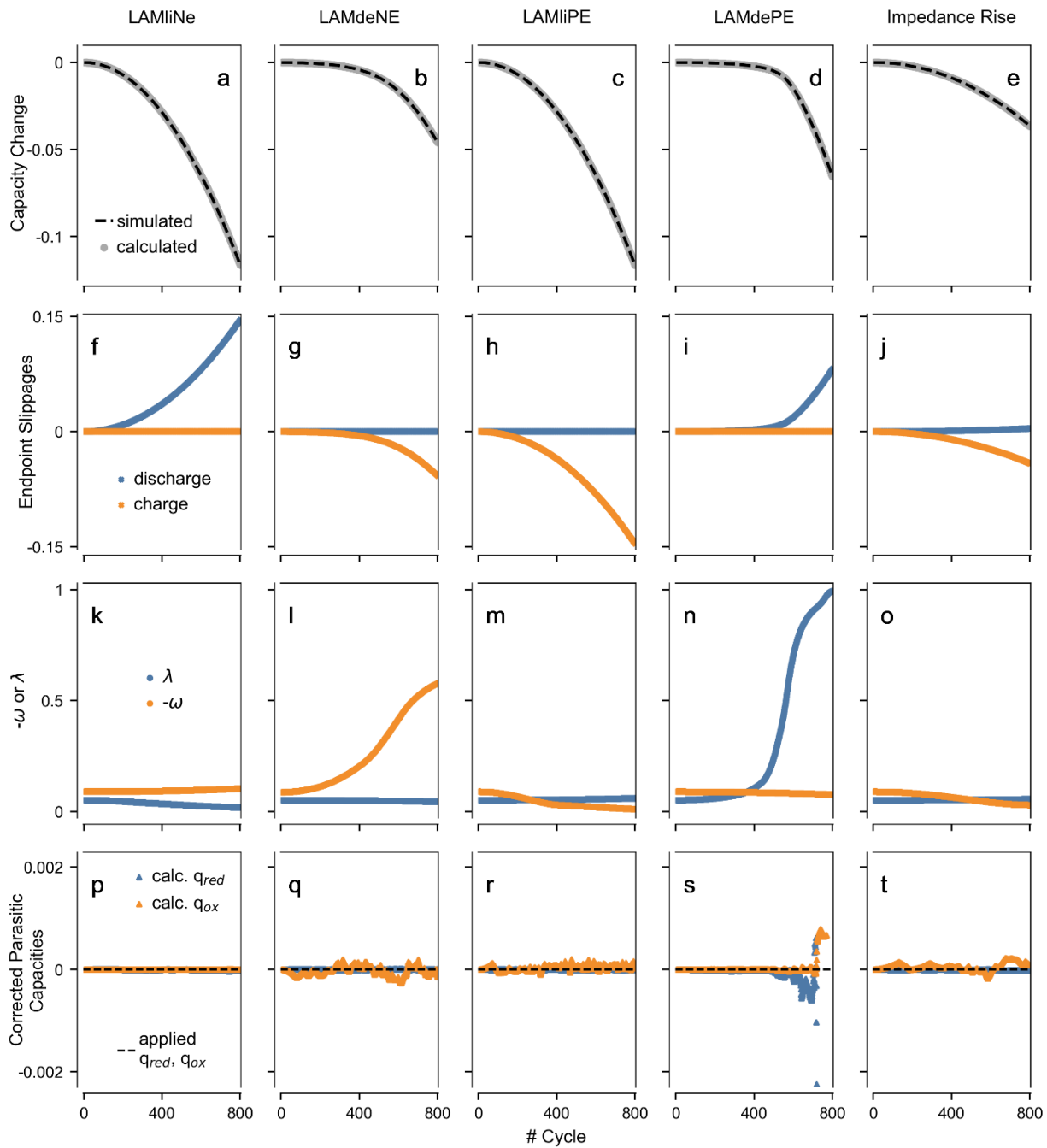

Figure 4. Simulated aging of a NMC532 vs graphite cell. a-e) cumulative capacity change; f-j) cumulative endpoint slippages; k-o) values for the parameters $\lambda$ and $\omega$; p-t) cumulative extent of reduction and oxidation side-reactions, both applied in the simulations and calculated using equations in Table 1. Each column corresponds to a single aging mode (see header) assumed to act on the cell according to Figure 3. The final extent of aging is the same as represented in Figure 2; refer to that figure for visual reference of the consequences of aging on voltage profiles. Capacity fade is shown in units of the initial PE SOC scale, so that it can be directly correlated with the amount of LAM in Figure 3. Note zoomed-in scale of bottom row. The equations in Table 1 enable the correct identification that, in all cases, the cell experienced negligible parasitic reduction and oxidation.

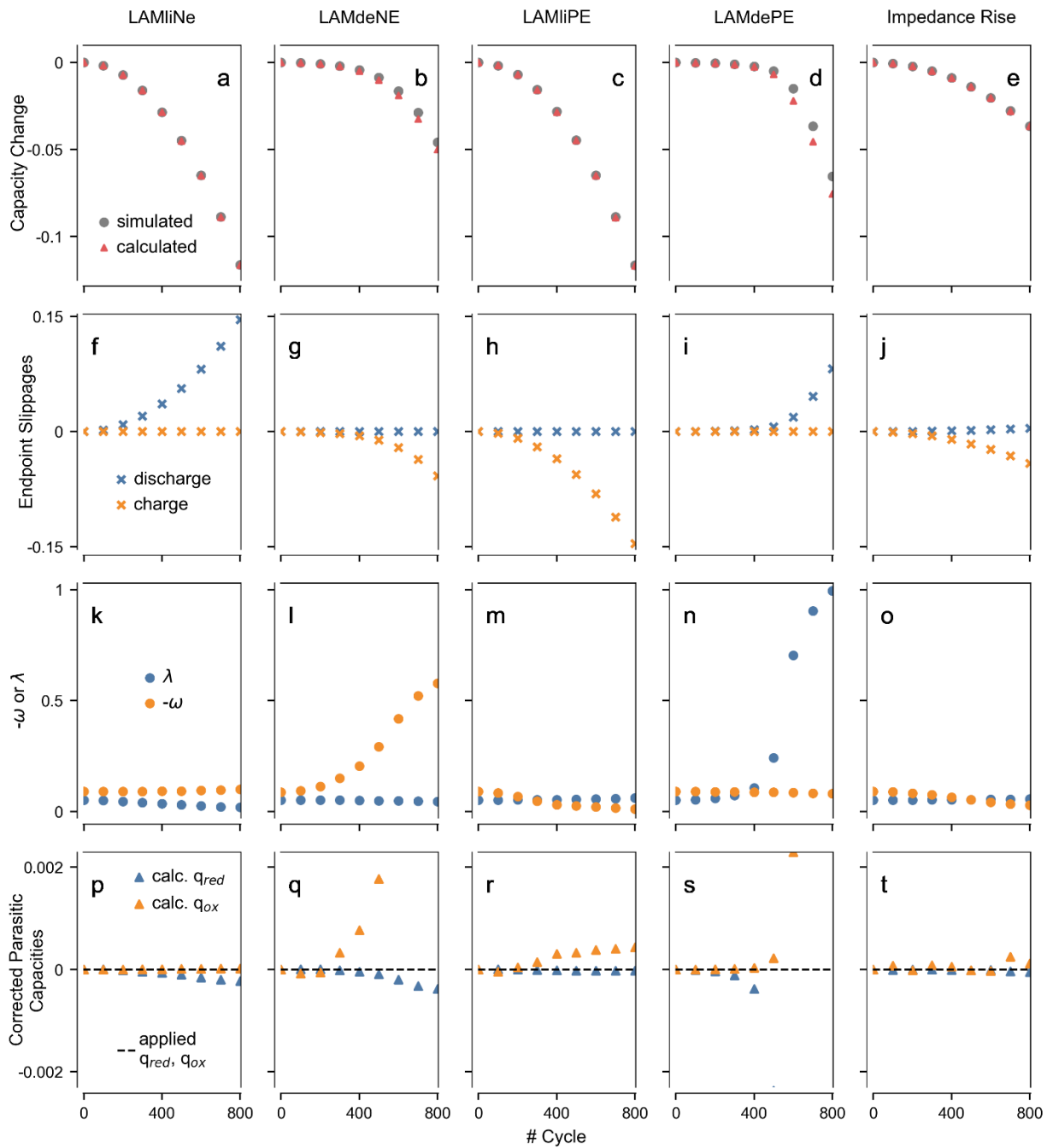

Figure 5. Like Figure 4 but considering one data point every 100 cycles for calculations. Errors appear for LAMdeNE and LAMdePE starting at the point where the electrode limiting the charge and discharge, respectively, changes. This change leads to large variations in $\lambda$ and $\omega$, rendering the equations in Table 1 less accurate. These extreme cases are relatively uncommon, and equations can be successfully used when probing occasional, non-consecutive cycles, such as reference performance tests.

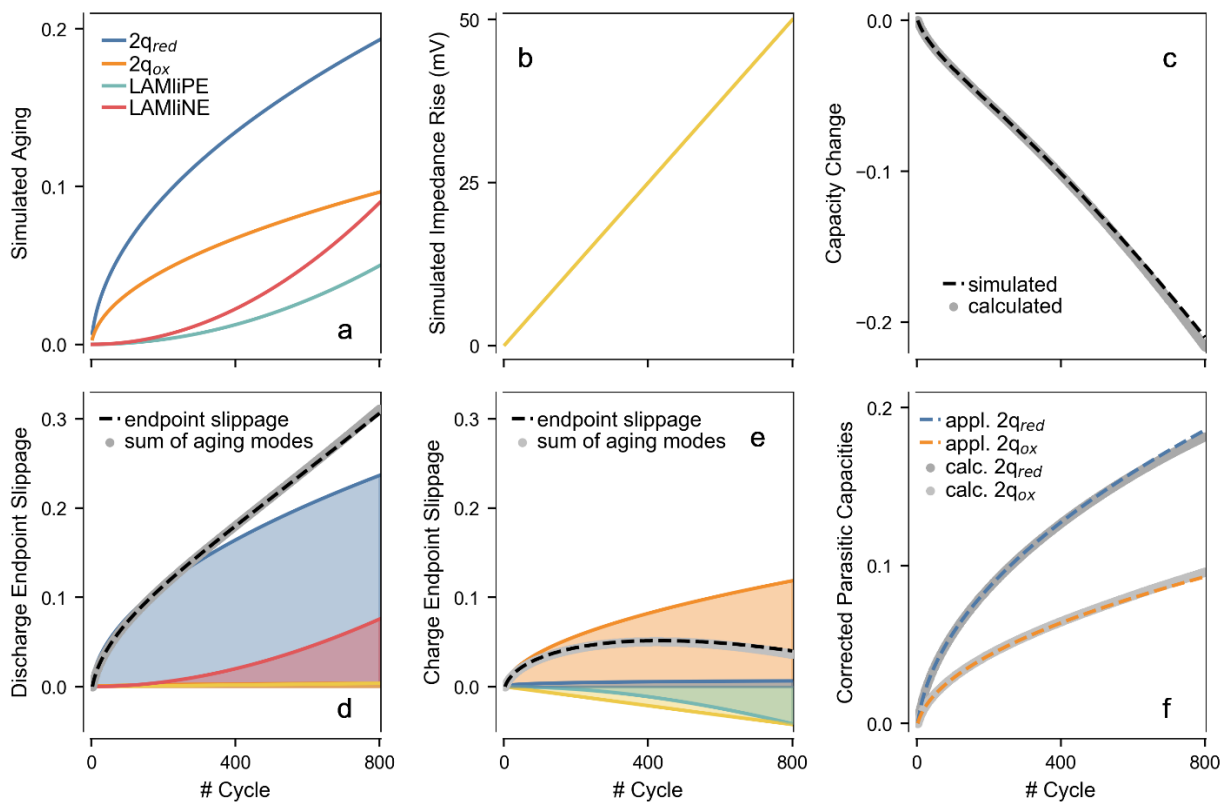

Figure 6. Multi-modal aging. a) Cumulative extent of LAM and side-reactions applied onto a NMC532 vs graphite cell, all in units of the initial PE SOC scale. b) Assumed cumulative increase in PE polarization caused by impedance rise. c) Cumulative capacity change. d) Cumulative discharge endpoint slippage. e) Cumulative charge endpoint slippage. f) Cumulative extent of reduction and oxidation side-reactions, both applied in simulations and calculated using equations in Table 1. The shaded areas in panels d and e highlight the contributions of individual modes to the slippages. Even in the presence of multiple aging modes, the equations in Table 1 can successfully identify the true extent of side-reactions experienced by the cell.

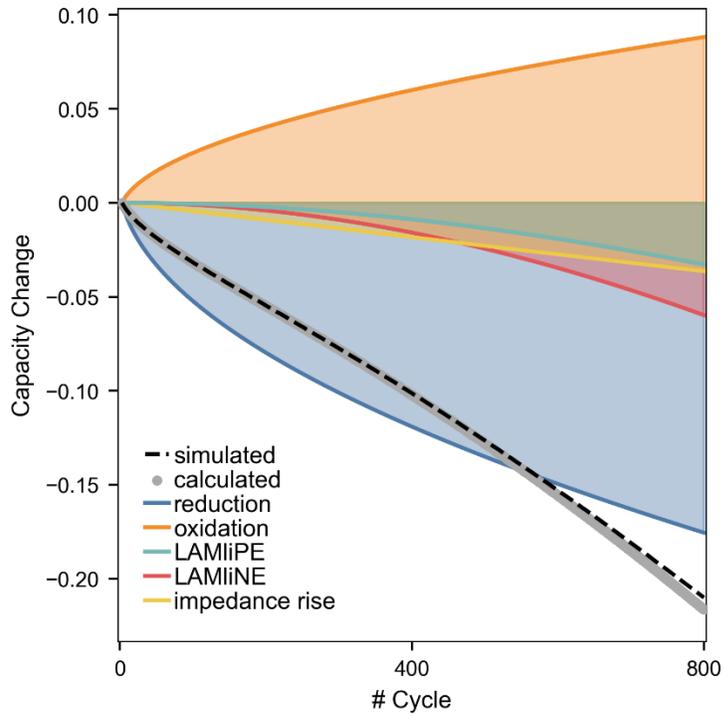

Figure 7. The cumulative capacity change of Figure 6c, but partitioned into the individual contributions of the various aging modes. Most of the change is driven by parasitic reduction and oxidation.

# Supplementary Material

# Endpoint slippage analysis in the presence of impedance rise and loss of active material

Marco-Tulio F. Rodrigues

Chemical Sciences and Engineering Division, Argonne National Laboratory, Lemont, IL, USA

**Contact:** marco@anl.gov

## Section S1. Definitions

Definitions:

- PE = positive electrode
- NE = negative electrode
- EOC = end of charge
- EOD = end of discharge
- SOC = state of charge
- $Q_c$ and $Q_d$ are the measurable full-cell charge and discharge capacities, respectively
- $Q_{BOC}$ is the amount of capacity that would be exchanged during the charge half-cycle in the absence of parasitic processes within that half-cycle (i.e., the ideal accessible Li$^+$ inventory at the beginning of charge)
- $Q_{BOD}$ is the amount of capacity that would be exchanged during the discharge half-cycle in the absence of parasitic processes within that half-cycle (i.e., the ideal accessible Li$^+$ inventory at the beginning of discharge)
- $\left.\frac{dU_{i,j}}{dq}\right|_k$ is the slope of the voltage profile of the electrode i (i = PE, NE for positive and negative electrodes, respectively) during the half-cycle j (j = c,d for charge and discharge, respectively) around the point k (k = EOC and EOD). It is assumed that the slope remains constant within the small variation in electrode SOC as the cell ages within a single cycle. (That is, the slope of the NE profile is assumed to be approximately the same at the end of two successive discharges despite the occurrence of a small shift in the electrode potential at the EOD.)

- $\left.\frac{dU_{i,j}}{dq}\right|_k$ is measured within the frame of reference in which other calculations are performed. For example, most of the analyses in the present manuscript are made with respect to the PE SOC scale, and hence electrodes need to be represented in that scale prior to computing the derivative. Similarly, when analyzing experimental data, voltage profiles for the electrodes must first be represented at the relevant scale

- $\left.\frac{dU_{i,c}}{dq}\right|_k = -\left.\frac{dU_{i,d}}{dq}\right|_k$, as we are comparing the electrode profiles during charge and discharge

- $\left.\frac{dU_{PE,c}}{dq}\right|_{EOC}$ and $\left.\frac{dU_{NE,d}}{dq}\right|_{EOD}$ are $> 0$; $\left.\frac{dU_{PE,d}}{dq}\right|_{EOD}$ and $\left.\frac{dU_{NE,c}}{dq}\right|_{EOC}$ are $< 0$

- Electrode potentials at the end of a half-cycle are referred to as "terminal potentials"

- $L_{j,k}$ is the LAM of type $j$ experienced between cycles $k-1$ and $k$, expressed in the SOC scale of the affected electrode

- $N_0$ is the ratio between the initial capacities of the NE and PE profiles used to simulate/fit cells

- $x$ is the Li$^+$ content of the NE (measured from 0 to 1)

- $y$ is the state of delithiation of the PE (measured from 0 to 1)

- $I\Delta R_k$ is the change in PE polarization under a current $I$ caused by the impedance rise $\Delta R_k$ experienced before the onset of cycle $k$

From our previous work (ref. S1), it has been shown that the slippage of endpoints between cycles $k-1$ and $k$ caused by aging mechanisms are given by

$$D_{slip,k} = D_k - D_{k-1} = Q_{c,k} - Q_{d,k} \tag{S1}$$

$$C_{slip,k} = C_k - C_{k-1} = Q_{c,k} - Q_{d,k-1} \tag{S2}$$

where $D_{slip}$ and $C_{slip}$ are the discharge and charge endpoint slippages, respectively, between cycles $k-1$ and $k$, and $D$ and $C$ are the endpoints at the indicated cycles. Knowing expressions relating $Q_c$, $Q_d$ and $Q_{d,k-1}$ to aging mechanisms, we can determine $D_{slip}$ and $C_{slip}$. We have already done that for the case of parasitic oxidation and reduction (ref. S1). Here, we extend this analysis to other aging modes.

From the expressions above, it also follows that the cycle-to-cycle capacity fade $Q_{loss}$ is given by

$$Q_{loss,k} = Q_{d,k-1} - Q_{d,k} = D_{slip,k} - C_{slip,k} \tag{S3}$$

## Section S2. Loss of active material (negative electrode) in the lithiated state (LAMliNE)

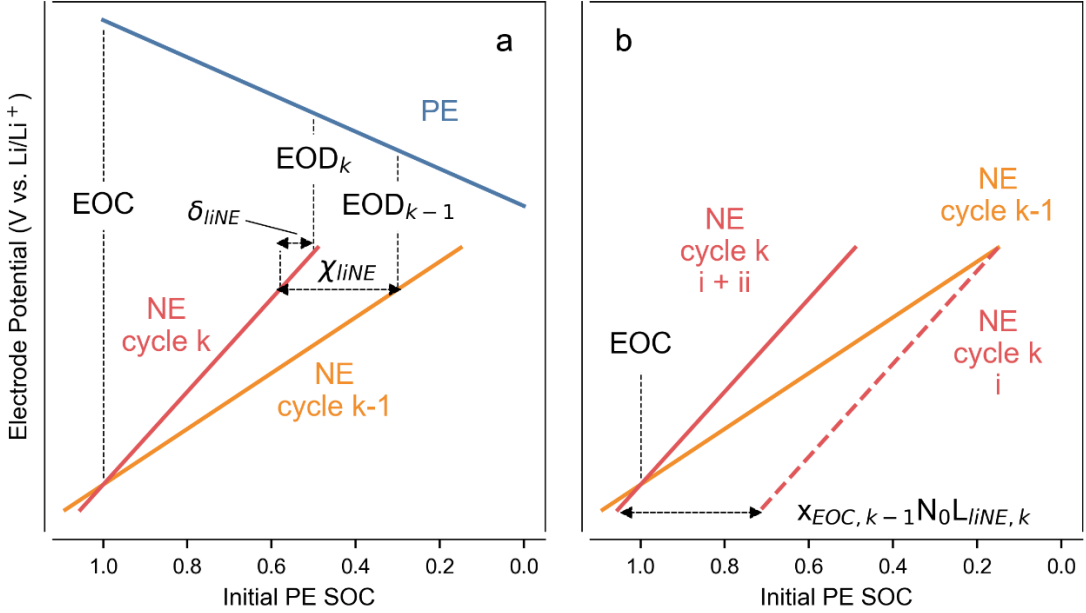

*Figure S1. a) Hypothetical voltage profiles for the discharge of a cell experiencing LAMliNE at the top of charge of cycle $k$. Quantities of interest for the derivation in this section are shown in the figure. b) Transformations of the NE profile captured by $\chi$, including the shrinkage of the profile (i) and offset by the indicated quantity (ii). The offset ensures that the EOC of the cell remains constant.*

LAMliNE involves the inactivation of negative electrode domains when the cell is in the charged state at an arbitrary cycle $k$ (Figure S1a). Disconnection of the lithiated NE particles will also lead to loss of the lithium ions within. Since LAMliNE acts on the charged cell, it does not have a direct effect on the preceding charge half-cycle itself, and thus

$$Q_c = Q_{BOC} \tag{S4}$$

Here, $Q_{BOC}$ denotes the ideal capacity achievable during the charging half-cycle in the absence of aging phenomena. The discharge capacity $Q_d$, however, will be affected by LAMliNE, and can be expressed as

$$Q_d = Q_{BOD} - \chi_{liNE} - \delta_{liNE} \tag{S5}$$

Here, $Q_{BOD}$ is the ideal discharge capacity in the absence of aging. The term $\chi_{liNE}$ accounts for the change in capacity due to "shrinkage" of the NE voltage profile. More specifically, we define $\chi_{liNE}$ as the difference in capacity between the NE profiles at cycles $k$ and $k-1$ at the potential experienced by the NE at the EOD of cycle $k-1$ (see Figure S1a). The term $\delta_{liNE}$ accounts for additional changes in cell capacity due to alterations in the terminal electrode potentials at the EOD (Figure S1a). Since LAMliNE affects how much the PE can be relithiated before the cell reaches the EOD, its occurrence may cause changes to the potentials experienced by the NE and PE at the end of discharge. In summary, the discharge capacity measured after LAMliNE is expressed as the "ideal" capacity in the absence of aging ($Q_{BOD}$), minus the shrinkage of the NE profile at a fixed EOD potential ($\chi_{liNE}$), minus any additional capacity exchange that is required to accommodate the changes in electrode EOD potential in the aged cell ($\delta_{liNE}$).

In keeping with the sign convention of our previous work, $\delta_{liNE}$ is positive when it involves additional Li$^+$ extraction from the PE and insertion into the NE, and negative if it involves the opposite process (that is, additional delithiation of the NE and relithiation of the PE). Combining these two terms ($\chi_{liNE}$ and $\delta_{liNE}$), we can fully describe the changes in cell capacity caused by shrinkage of the NE voltage profile, loss of the Li$^+$ within disconnected NE particles and changes in the EOD potentials of both electrodes.

To find $\chi_{liNE}$, we consider that we are looking for the change in capacity value (here, in unaged PE SOC scale), as a result of LAM, for the point in the NE profile equal to the NE EOD potential at the preceding cycle (that is, prior to the occurrence of LAMliNE). Figure S1b shows that this change can be described by two components: i) a shrinkage of the NE voltage profile; and ii) a shift of the NE profile towards higher values in the PE SOC scale. The former captures the LAM per se, while the latter accounts for the Li$^+$ that is lost within the disconnected NE domains.

The amount of shrinkage of the NE voltage profile increases linearly with the Li$^+$ content $x$ within the NE, being zero for $x = 0$ and equal to $N_0 L_{liNE,k}$ for $x = 1$ (Figure 1b, dashed red trace). Here, multiplication by $N_0$ effectively converts $L_{liNE,k}$ from the NE to the PE SOC scale. At the potential matching the EOD of cycle $k - 1$, the shrinkage is given by $x_{EOD,k-1} N_0 L_{liNE,k}$. For the second term, the additional offset caused by the Li$^+$ loss can be described by $x_{EOC,k} N_0 L_{liNE,k}$, where $x_{EOC,k}$ indicates the Li$^+$ content of the NE at the moment of disconnection. Considering that the first term moves a given point of the NE voltage profile towards lower SOCs (in the PE SOC scale, Figure S1b), and the second moves it towards higher, we have

$$\chi_{liNE} = x_{EOC,k} N_0 L_{liNE,k} - x_{EOD,k-1} N_0 L_{liNE,k} = N_0 L_{liNE,k}(x_{EOC,k} - x_{EOD,k-1}) \quad (S6)$$

Note that LAMliNE leads to isolation of a fraction of the NE but has no effect on the Li$^+$ content of either the isolated or the remaining active domains. Consequently, it does not change the EOC (or the charge endpoint slippage, as we will show below), making $x_{EOC,k} = x_{EOC,k-1}$ and thus

$$\chi_{liNE} = N_0 L_{liNE,k}(x_{EOC,k-1} - x_{EOD,k-1}) \qquad (S7)$$

To calculate $\delta_{d,liNE}$, we must consider that changes to the EOD potentials experienced by the PE and the NE before and after LAMliNE must be equal. In other words, an increase in the PE EOD potential would cause an identical increase in the NE EOD potential if a constant lower cutoff voltage is assumed for the cell. In accordance with our prior work, we assume that the slopes of the voltage profile at the EOD are such that $\left.\frac{dU_{NE,d}}{dq}\right|_{EOD} > 0$ and $\left.\frac{dU_{PE,d}}{dq}\right|_{EOD} < 0$. Additional considerations:

- LAMliNE will generally move the EOD towards higher PE potentials, which will also force the EOD to occur at proportionally higher NE potentials. Since this leads to additional delithiation of the NE, $\delta_{liNE}$ is negative.
- The change in the terminal PE potential directly caused by shrinkage of the NE voltage profile is given by $-\chi_{liNE} \left.\frac{dU_{PE,d}}{dq}\right|_{EOD}$, where the negative sign is needed to turn this quantity into a positive number, as $\chi_{liNE} > 0$.
- The increase in the terminal NE potential is given by $-\delta_{liNE} \left.\frac{dU_{NE,d}}{dq}\right|_{EOD}$, where the negative sign is needed to turn this quantity into a positive number.
- The partial decrease in terminal PE potential caused by the additional NE delithiation is given by $-\delta_{liNE} \left.\frac{dU_{PE,d}}{dq}\right|_{EOD}$, where the negative sign is needed to turn this quantity into a negative number.

Hence,

$$-\delta_{liNE} \left.\frac{dU_{NE,d}}{dq}\right|_{EOD} = -\delta_{liNE} \left.\frac{dU_{PE,d}}{dq}\right|_{EOD} - \chi_{liNE} \left.\frac{dU_{PE,d}}{dq}\right|_{EOD} \quad (S8)$$

$$\delta_{liNE} = \frac{-\chi_{liNE} \left.\frac{dU_{PE,d}}{dq}\right|_{EOD}}{\left(\left.\frac{dU_{PE,d}}{dq}\right|_{EOD} - \left.\frac{dU_{NE,d}}{dq}\right|_{EOD}\right)} \quad (S9)$$

Retrieving the definition from our prior works (refs. S1-S3) that

$$\lambda = \frac{\left.\frac{dU_{PE,d}}{dq}\right|_{EOD}}{\left(\left.\frac{dU_{PE,d}}{dq}\right|_{EOD} - \left.\frac{dU_{NE,d}}{dq}\right|_{EOD}\right)} \quad (S10)$$

we have that

$$\delta_{liNE} = -\chi_{liNE} \lambda \quad (S11)$$

Substituting equations S7 and S11 into eq. S5,

$$Q_d = Q_{BOD} - \chi_{liNE} + \chi_{liNE}\lambda = Q_{BOD} - \chi_{liNE}(1 - \lambda) \quad (S12)$$

$$Q_d = Q_{BOD} - N_0 L_{liNE,k}(x_{EOC,k-1} - x_{EOD,k-1})(1 - \lambda) \quad (S13)$$

Since the LAMliNE happening on cycle $k$ has no direct effect on the charge capacity at that cycle,

$$Q_{BOD} = Q_c = Q_{BOC} \quad (S14)$$

from which it follows that

$$Q_d = Q_{BOC} - N_0 L_{liNE,k}(x_{EOC,k-1} - x_{EOD,k-1})(1 - \lambda) \tag{S15}$$

It is also true that the discharge capacity of cycle $k - 1$ is such that

$$Q_{d,k-1} = Q_c = Q_{BOC} \tag{S16}$$

and hence

$$D_{slip,k} = Q_c - Q_d = N_0 L_{liNE,k}(x_{EOC,k-1} - x_{EOD,k-1})(1 - \lambda) \tag{S17}$$

$$C_{slip,k} = Q_c - Q_{d,k-1} = Q_{BOC} - Q_{BOC} = 0 \tag{S18}$$

$$Q_{loss,k} = D_{slip,k} - C_{slip,k} = N_0 L_{liNE,k}(x_{EOC,k-1} - x_{EOD,k-1})(1 - \lambda) \tag{S19}$$

For many graphite-based cells operating between 0 and 1 SOC, we can approximate $\lambda \approx \omega \approx 0$ to get

$$D_{slip,k} = N_0 L_{liNE,k}(x_{EOC,k-1} - x_{EOD,k-1}) \tag{S20}$$

$$Q_{loss,k} = D_{slip,k} - C_{slip,k} = N_0 L_{liNE,k}(x_{EOC,k-1} - x_{EOD,k-1}) \tag{S21}$$

## Section S3. Loss of active material (negative electrode) in the delithiated state (LAMdeNE)

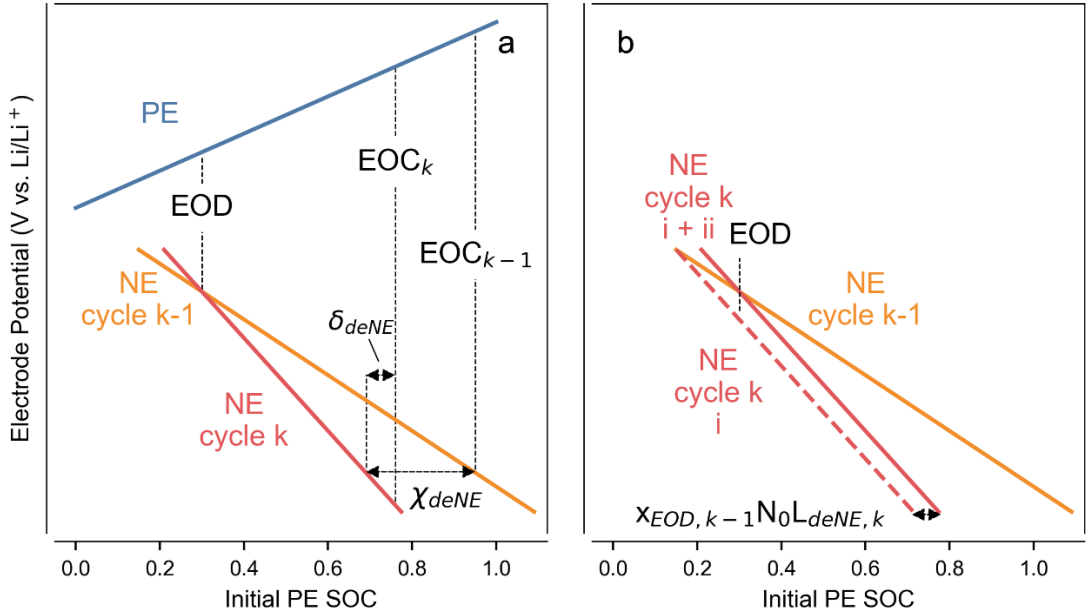

*Figure S2.* a) Hypothetical voltage profiles for the charge of a cell experiencing LAMdeNE before the charge of cycle $k$. Quantities of interest for the derivation in this section are shown in the figure. b) Transformations of the NE profile captured by $\chi$, including the shrinkage of the profile (i) and offset by the indicated quantity (ii). The offset ensures that the EOD of the cell remains constant.

LAMdeNE involves the inactivation of negative electrode domains when the cell is in the discharged state at an arbitrary cycle $k$ (Figure S2a). Note that disconnection of the delithiated NE particles will still lead to loss of some lithium ions within, as the NE is often not fully delithiated at the EOD. LAMdeNE can be considered to occur prior to cell charging and will affect $Q_c$ according to

$$Q_c = Q_{BOC} + \chi_{deNE} + \delta_{deNE} \tag{S22}$$

Here, definitions are like the used in the previous section, with $\chi_{deNE}$ capturing capacity loss due to shrinking of the NE voltage profile and due to the Li$^+$ trapped within the disconnected particles, and $\delta_{deNE}$ describing additional capacity exchanged as the aged cell settles at a new combination of electrode potentials at the EOC.

Applying the same logic as before, the NE voltage profile at cycle $k$ (that is, after LAMdeNE) will meet the EOC NE potential of cycle $k-1$ at capacities differing by $x_{EOC,k-1} N_0 L_{deNE,k}$. A small offset may be applied over the NE voltage profile due to Li$^+$ loss at the (partially) delithiated disconnected particles, which can be described by $x_{EOD,k-1} N_0 L_{deNE,k}$ (Figure S2b). Note that the former term moves the profile towards lower SOCs, while the latter moves it towards higher. Combining these two contributions, we get

$$\chi_{deNE} = x_{EOD,k-1} N_0 L_{deNE,k} - x_{EOC,k-1} N_0 L_{deNE,k}$$
$$= -N_0 L_{deNE,k} (x_{EOC,k-1} - x_{EOD,k-1}) \tag{S23}$$

and $\chi_{deNE} < 0$. For $\delta_{deNE}$, we consider that establishing a new EOC requires equal changes to terminal potentials in the PE and NE, which can be written as

$$\delta_{deNE} \left. \frac{dU_{NE,c}}{dq} \right|_{EOC} = \delta_{deNE} \left. \frac{dU_{PE,c}}{dq} \right|_{EOC} + \chi_{deNE} \left. \frac{dU_{PE,c}}{dq} \right|_{EOC} \tag{S24}$$

Here, $\left. \frac{dU_{NE,c}}{dq} \right|_{EOC} < 0$, $\left. \frac{dU_{PE,c}}{dq} \right|_{EOC} > 0$ and $\delta_{deNE} > 0$. It follows that

$$\delta_{deNE} = \frac{-\chi_{deNE} \left.\frac{dU_{PE,c}}{dq}\right|_{EOC}}{\left.\frac{dU_{PE,c}}{dq}\right|_{EOC} - \left.\frac{dU_{NE,c}}{dq}\right|_{EOC}} \tag{S25}$$

Borrowing from our previous works (refs. S1-S3) that

$$\frac{\left.\frac{dU_{NE,c}}{dq}\right|_{EOC}}{\left.\frac{dU_{PE,c}}{dq}\right|_{EOC} - \left.\frac{dU_{NE,c}}{dq}\right|_{EOC}} = \omega \tag{S26}$$

$$\frac{\left.\frac{dU_{PE,c}}{dq}\right|_{EOC}}{\left.\frac{dU_{PE,c}}{dq}\right|_{EOC} - \left.\frac{dU_{NE,c}}{dq}\right|_{EOC}} = 1 + \omega \tag{S27}$$

We can rewrite equation S25 as

$$\delta_{deNE} = -\chi_{deNE}(1 + \omega) \tag{S28}$$

Finally, that makes

$$Q_c = Q_{BOC} + \chi_{deNE} - \chi_{deNE}(1 + \omega) = Q_{BOC} - \omega\chi_{deNE} \tag{S29}$$

$$Q_c = Q_{BOC} + N_0 L_{deNE,k}\omega(x_{EOC,k-1} - x_{EOD,k-1}) \tag{S30}$$

Since LAMdeNE operates prior to $Q_c$, we have that

$$Q_d = Q_c \tag{S31}$$

$$Q_{d,k-1} = Q_{BOC} \tag{S32}$$

With the expressions above, we can write

$$D_{slip,k} = Q_c - Q_d = Q_c - Q_c = 0 \tag{S33}$$

$$\begin{aligned} C_{slip,k} = Q_c - Q_{d,k-1} &= Q_{BOC} + N_0 L_{deNE,k}\omega(x_{EOC,k-1} - x_{EOD,k-1}) - Q_{BOC} \\ &= N_0 L_{deNE,k}\omega(x_{EOC,k-1} - x_{EOD,k-1}) \end{aligned} \tag{S34}$$

$$Q_{loss,k} = D_{slip,k} - C_{slip,k} = -N_0 L_{deNE,k}\omega(x_{EOC,k-1} - x_{EOD,k-1}) \tag{S35}$$

Note that $-1 \leq \omega \leq 0$, and so equation S34 can lead to slippage of the charge endpoint towards lower values.

For many graphite-based cells operating between 0 and 1 SOC, we can approximate $\lambda \approx \omega \approx 0$ to get

$$C_{slip,k} = 0 \tag{S36}$$

$$Q_{loss,k} = 0 \tag{S37}$$

## Section S4. Loss of active material (positive electrode) in the lithiated state (LAMliPE)

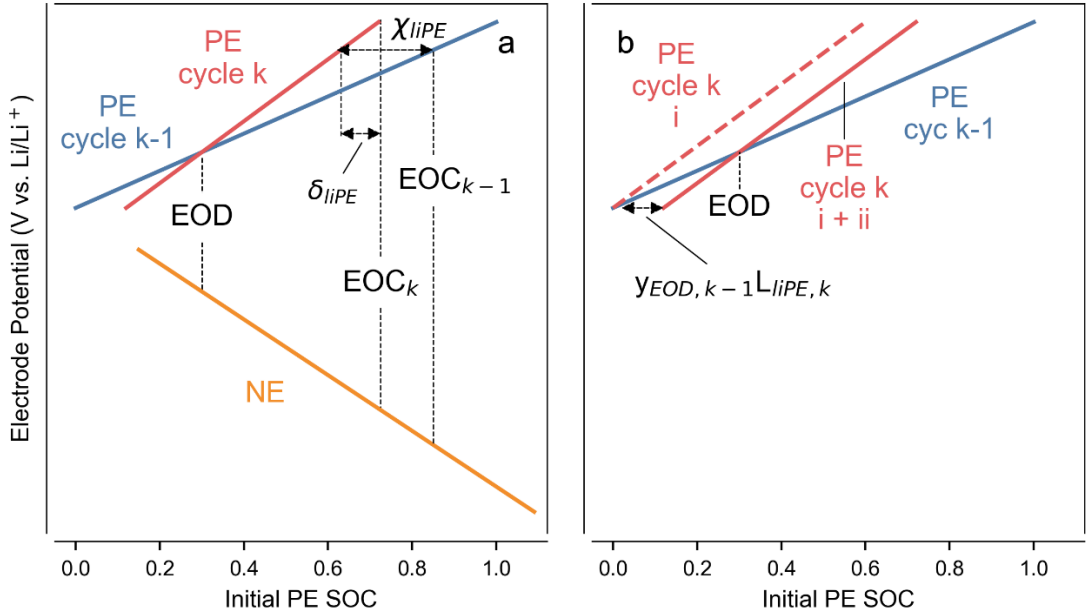

*Figure S3. a) Hypothetical voltage profiles for the charge of a cell experiencing LAMliPE before the charge of cycle $k$. Quantities of interest for the derivation in this section are shown in the figure. b) Transformations of the PE profile captured by $\chi$, including the shrinkage of the profile (i) and offset by the indicated quantity (ii). The offset ensures that the EOD of the cell remains constant.*

Similar to LAMdeNE, we can consider that LAMliPE occurs in between discharge $k-1$ and charge $k$, and that the charge capacity of the cell can be written as

$$Q_c = Q_{BOC} + \chi_{liPE} + \delta_{liPE} \tag{S38}$$

Following the same logic used above, the term $\chi_{liPE}$ (Figure S3) contains information about: i) the shrinkage of the PE profile; and ii) a slight shift of the PE profile due to the eventual trapping of $Li^+$ within the disconnected particles, as the positive electrode is unlikely to be fully relithiated at the moment of material loss (Figure S3b). Since all equations so far have been written in the frame of reference of the PE SOC, we can use the PE state of delithiation $y$ to write

$$\chi_{liPE} = y_{EOD,k-1} L_{liPE,k} - y_{EOC,k-1} L_{liPE,k} = -L_{liPE,k}(y_{EOC,k-1} - y_{EOD,k-1}) \tag{S39}$$

The capacity $\delta_{liPE}$ exchanged as the electrodes settle at a new cell EOC state is obtained from the balance of changes to electrode potentials, according to

$$\delta_{liPE} \left.\frac{dU_{NE,c}}{dq}\right|_{EOC} + \chi_{liPE} \left.\frac{dU_{NE,c}}{dq}\right|_{EOC} = \delta_{liPE} \left.\frac{dU_{PE,c}}{dq}\right|_{EOC} \tag{S40}$$

Here, $\left.\frac{dU_{NE,c}}{dq}\right|_{EOC} < 0$, $\left.\frac{dU_{PE,c}}{dq}\right|_{EOC} > 0$ and $\delta_{deNE} > 0$. It follows that

$$\delta_{liPE} = \frac{\chi_{liPE} \left.\frac{dU_{NE,c}}{dq}\right|_{EOC}}{\left.\frac{dU_{PE,c}}{dq}\right|_{EOC} - \left.\frac{dU_{NE,c}}{dq}\right|_{EOC}} = \chi_{liPE}\omega \tag{S41}$$

$$Q_c = Q_{BOC} + \chi_{liPE} + \delta_{liPE} = Q_{BOC} + \chi_{liPE} + \chi_{liPE}\omega = Q_{BOC} + \chi_{liPE}(1+\omega) \tag{S42}$$

$$Q_c = Q_{BOC} - L_{liPE,k}(y_{EOC,k-1} - y_{EOD,k-1})(1+\omega) \tag{S43}$$

$$Q_d = Q_c \tag{S44}$$

$$Q_{d,k-1} = Q_{BOC} \tag{S45}$$

$$D_{slip,k} = Q_c - Q_d = Q_c - Q_c = 0 \tag{S46}$$

$$C_{slip,k} = Q_c - Q_{d,k-1} = Q_{BOC} - L_{liPE,k}(y_{EOC,k-1} - y_{EOD,k-1})(1+\omega) - Q_{BOC}$$
$$= -L_{liPE,k}(y_{EOC,k-1} - y_{EOD,k-1})(1+\omega) \tag{S47}$$

$$Q_{loss,k} = D_{slip,k} - C_{slip,k} = L_{liPE,k}(y_{EOC,k-1} - y_{EOD,k-1})(1 + \omega) \tag{S48}$$

For many graphite-based cells operating between 0 and 1 SOC, we can approximate $\lambda \approx \omega \approx 0$ to get

$$C_{slip,k} = -L_{liPE,k}(y_{EOC,k-1} - y_{EOD,k-1}) \tag{S49}$$

$$Q_{loss,k} = L_{liPE,k}(y_{EOC,k-1} - y_{EOD,k-1}) \tag{S50}$$

# Section S5. Loss of active material (positive electrode) in the delithiated state (LAMdePE)

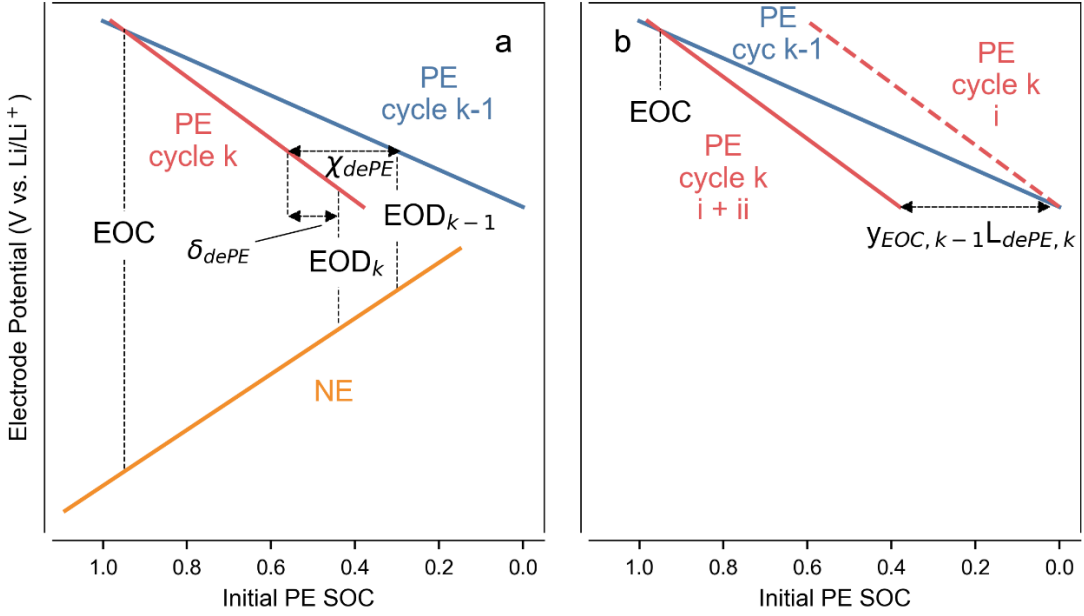

**Figure S4.** *a) Hypothetical voltage profiles for the discharge of a cell experiencing LAMdePE at the top of charge of cycle k. Quantities of interest for the derivation in this section are shown in the figure. b) Transformations of the PE profile captured by χ, including the shrinkage of the profile (i) and offset by the indicated quantity (ii). The offset ensures that the EOC of the cell remains constant.*

Following the steps used above, we can write

$$Q_d = Q_{BOD} - \chi_{dePE} - \delta_{dePE} \tag{S51}$$

with the term $\chi_{dePE}$ (Figure S4) defined as

$$\chi_{dePE} = y_{EOC,k-1} L_{dePE,k} - y_{EOD,k-1} L_{dePE,k} = L_{dePE,k}(y_{EOC,k-1} - y_{EOD,k-1}) \tag{S52}$$

$\delta_{dePE}$ can once again be determined based on the balance of changes to the electrode potentials when establishing a new EOD, as

$$-\delta_{dePE} \left.\frac{dU_{NE,d}}{dq}\right|_{EOD} - \chi_{dePE} \left.\frac{dU_{NE,d}}{dq}\right|_{EOD} = -\delta_{dePE} \left.\frac{dU_{PE,d}}{dq}\right|_{EOD} \tag{S53}$$

with $\left.\frac{dU_{NE,d}}{dq}\right|_{EOD} > 0$, $\left.\frac{dU_{PE,d}}{dq}\right|_{EOD} < 0$ and $\delta_{dePE} < 0$. It follows that

$$\delta_{dePE} = \frac{\chi_{dePE} \left.\frac{dU_{NE,d}}{dq}\right|_{EOD}}{\left.\frac{dU_{PE,d}}{dq}\right|_{EOD} - \left.\frac{dU_{NE,d}}{dq}\right|_{EOD}} = \chi_{dePE}(\lambda - 1) \tag{S54}$$

$$Q_d = Q_{BOD} - \chi_{dePE} - \delta_{dePE} = Q_{BOD} - \chi_{dePE} - \chi_{dePE}(\lambda - 1) = Q_{BOD} - \lambda\chi_{dePE} \tag{S55}$$

$$Q_d = Q_{BOD} - L_{dePE,k}\lambda(y_{EOC,k-1} - y_{EOD,k-1}) \tag{S56}$$

$$Q_c = Q_{BOC} = Q_{BOD} \tag{S57}$$

$$Q_{d,k-1} = Q_c \tag{S58}$$

$$D_{slip,k} = Q_c - Q_d = Q_{BOD} - Q_{BOD} + L_{dePE,k}\lambda(y_{EOC,k-1} - y_{EOD,k-1})$$
$$= L_{dePE,k}\lambda(y_{EOC,k-1} - y_{EOD,k-1}) \tag{S59}$$

$$C_{slip,k} = Q_c - Q_{d,k-1} = Q_c - Q_c = 0 \tag{S60}$$

$$Q_{loss,k} = D_{slip,k} - C_{slip,k} = L_{dePE,k}\lambda(y_{EOC,k-1} - y_{EOD,k-1}) \tag{S61}$$

For many graphite-based cells operating between 0 and 1 SOC, we can approximate $\lambda \approx \omega \approx 0$ to get

$$D_{slip,k} = 0 \tag{S62}$$

$$Q_{loss,k} = 0 \tag{S63}$$

That is, LAMdePE will not have any measurable effect on the cell for as long as discharge is strongly limited by the NE.

## Section S6. Impedance rise

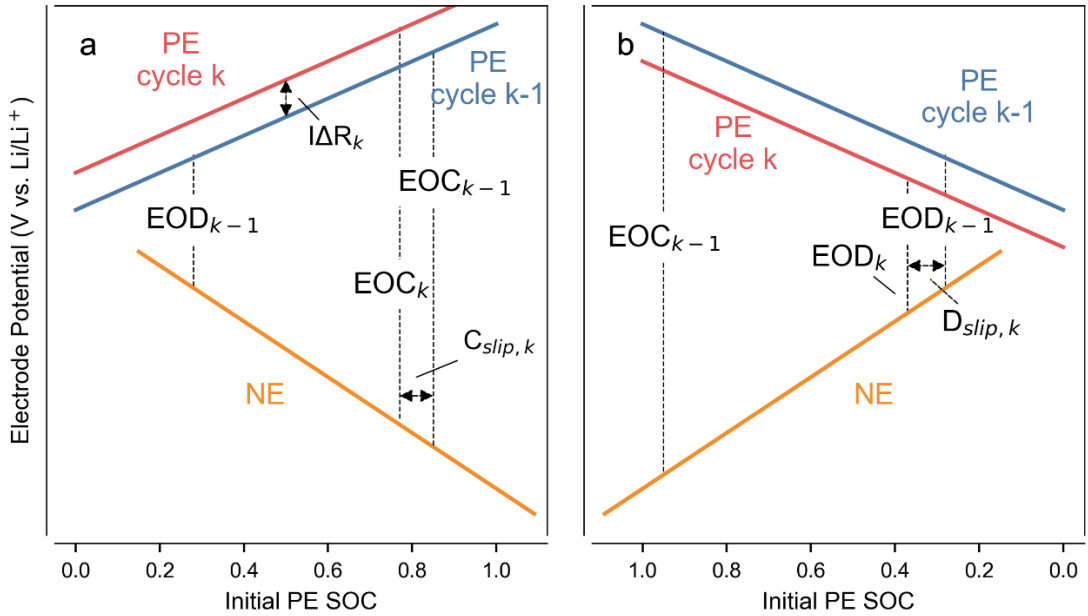

***Figure S5.*** *Hypothetical voltage profiles for the charge (a) and discharge (b) of a cell experiencing impedance rise before the charge of cycle k. Quantities of interest for the derivation in this section are shown in the figure.*

Consider that impedance rise affecting cycle $k$ happens right before the onset of that cycle. Experimentally, that is equivalent to measuring the impedance rise accrued in between two reference performance tests. As a first approximation, impedance rise at a given electrode will increase the polarization of the voltage profile, shifting it towards higher potentials during charge (Figure S5a) and to lower during discharge (Figure S5b). Since the voltage profile for the full-cell is the point-by-point difference between the voltage profiles for PE and NE, the gain in polarization experienced by the cell due to impedance rise is the same regardless of how it is split between the two electrodes. For simplicity, let us consider that all impedance rise happens at the PE. If the impedance rises by $\Delta R_k$, the increase in polarization will be given by $I\Delta R_k$, where $I$ is the current applied to the cell. Here, we use the quantity $I\Delta R_k$ to represent impedance rise.

An increase in PE polarization will cause the EOC point to shift towards a lower PE SOC (Figure S5a). This shift will increase the terminal potentials for both PE and NE, and decrease the charge endpoint by $C_{slip}$; we can infer from Figure S5a that $C_{slip} < 0$. Considering that the cell voltage cutoff that defines the EOC remains the same, the changes in PE and NE potentials must be equal, and so

$$I\Delta R_k + C_{slip,k} \frac{dU_{PE,c}}{dq}\bigg|_{EOC} = C_{slip,k} \frac{dU_{NE,c}}{dq}\bigg|_{EOC} \tag{S64}$$

$$C_{slip,k} = \frac{-I\Delta R_k}{\left(\frac{dU_{PE,c}}{dq}\bigg|_{EOC} - \frac{dU_{NE,c}}{dq}\bigg|_{EOC}\right)} = \frac{-I\Delta R_k}{\frac{dU_{cell,c}}{dq}\bigg|_{EOC}} \tag{S65}$$

A similar process will occur during cell discharge, with $D_{slip} > 0$ (Figure S5b). Here, the PE voltage profile will be depressed by $-I\Delta R_k$ (assuming that the current is always positive), and we have

$$-I\Delta R_k - D_{slip,k} \frac{dU_{PE,d}}{dq}\bigg|_{EOD} = -D_{slip,k} \frac{dU_{NE,d}}{dq}\bigg|_{EOD} \tag{S66}$$

$$D_{slip,k} = \frac{-I\Delta R_k}{\left(\frac{dU_{PE,d}}{dq}\bigg|_{EOD} - \frac{dU_{NE,d}}{dq}\bigg|_{EOD}\right)} = \frac{-I\Delta R_k}{\frac{dU_{cell,d}}{dq}\bigg|_{EOD}} \tag{S67}$$

$$Q_{loss,k} = D_{slip,k} - C_{slip,k} = \frac{-I\Delta R_k}{\left.\frac{dU_{cell,d}}{dq}\right|_{EOD}} + \frac{I\Delta R_k}{\left.\frac{dU_{cell,c}}{dq}\right|_{EOC}}$$

$$= I\Delta R_k \left( \frac{1}{\left.\frac{dU_{cell,c}}{dq}\right|_{EOC}} - \frac{1}{\left.\frac{dU_{cell,d}}{dq}\right|_{EOD}} \right) \quad (S68)$$

For certain systems operating between 0 and 1 SOC, like LFP vs graphite, the slope of the cell profile is large at both the EOC and EOD. In this case, we have

$$C_{slip,k} = 0 \quad (S69)$$

$$D_{slip,k} = 0 \quad (S70)$$

$$Q_{loss,k} = 0 \quad (S71)$$

## Section S7. General expressions for the endpoint slippages caused by LAM

Calculation of $\chi_{liNE}$ in Section S2 relies on the assumption that the EOC is unaffected by LAMliNE and remains static (Figure S1a). When multiple aging modes are acting on the cell, however, the EOC will likely vary from cycle to cycle, causing this assumption of invariance to be unrealistic and affecting the accuracy of the expression therein. Similar issues will arise for all other types of LAM. Here, we show how more general expressions can be found for LAMliNE. For brevity, expressions for the other modes are provided without derivation.

In Section S2 we define $\chi_{liNE}$ as the difference in capacity between the NE profiles at arbitrary cycles $k-1$ and $k$ at the potential exhibited by the NE at the EOD of cycle $k-1$. The coordinates $NE$ of the negative electrode along the PE SOC scale during discharge $k$ can be represented as

$$NE_k = xN_0(1 - L_{liNE,cum,k}) + x_{EOC,k}N_0 L_{liNE,cum,k} + \alpha \quad (S72)$$

where $x$ represents a given data point for Li$^+$ content, $L_{liNE,cum,k}$ is the total LAMliNE experienced by the electrode until cycle $k$ (inclusive), and $\alpha$ embodies the effects of any additional aging experienced by the cell (including the initial offset after formation cycles). The term $xN_0(1 - L_{liNE,cum,k})$ "shrinks" the voltage profile according to the Li$^+$ content $x$, while the term $x_{EOC,k}N_0 L_{liNE,cum,k}$ applies an offset to the NE profile due to trapping of the Li$^+$ within during LAM (see Figure S1b). Since we are interested in the value of $NE_k$ at the NE potential at the EOD of cycle $k-1$, the relevant equation is

$$NE_{EOD-1,k} = x_{EOD,k-1}N_0(1 - L_{liNE,cum,k}) + x_{EOC,k}N_0 L_{liNE,cum,k} + \alpha \quad (S73)$$

From here, we have that

$$\chi_{liNE} = NE_{EOD-1,k} - NE_{EOD-1,k-1}$$

$$= x_{EOD,k-1}N_0(1 - L_{liNE,cum,k}) + x_{EOC,k}N_0 L_{liNE,cum,k}$$

$$- x_{EOD,k-1}N_0(1 - L_{liNE,cum,k-1}) - x_{EOC,k-1}N_0 L_{liNE,cum,k-1}$$

$$= N_0[x_{EOD,k-1} - x_{EOD,k-1}L_{liNE,cum,k} + x_{EOC,k}L_{liNE,cum,k} - x_{EOD,k-1} \quad \text{(S74)}$$

$$+ x_{EOD,k-1}L_{liNE,cum,k-1} - x_{EOC,k-1}L_{liNE,cum,k-1}]$$

$$= N_0[-x_{EOD,k-1}(L_{liNE,cum,k} - L_{liNE,cum,k-1}) + x_{EOC,k}L_{liNE,cum,k}$$

$$- x_{EOC,k-1}L_{liNE,cum,k-1}]$$

But

$$L_{liNE,cum,k} - L_{liNE,cum,k-1} = L_{liNE,k} \quad \text{(S75)}$$

making

$$\chi_{liNE} = N_0[-x_{EOD,k-1}L_{liNE,k} + x_{EOC,k}L_{liNE,cum,k} - x_{EOC,k-1}(L_{liNE,cum,k} - L_{liNE,k})] \quad \text{(S76)}$$

and thus

$$\chi_{liNE} = N_0[(x_{EOC,k-1} - x_{EOD,k-1})L_{liNE,k} + (x_{EOC,k} - x_{EOC,k-1})L_{liNE,cum,k}] \quad \text{(S77)}$$

If $x_{EOC,k} = x_{EOC,k-1}$ (like assumed in Figure S1), equation S77 becomes equation S7. We can use this modified $\chi_{liNE}$ to find that

$$D_{slip,k} = N_0[(x_{EOC,k-1} - x_{EOD,k-1})L_{liNE,k} + (x_{EOC,k} - x_{EOC,k-1})L_{liNE,cum,k}](1 - \lambda) \tag{S78}$$

$$Q_{slip,k} = N_0[(x_{EOC,k-1} - x_{EOD,k-1})L_{liNE,k} + (x_{EOC,k} - x_{EOC,k-1})L_{liNE,cum,k}](1 - \lambda) \tag{S79}$$

Alternatively, we can maintain the LAM terms as in equation S74 to get

$$\begin{aligned}\chi_{liNE} &= N_0[-x_{EOD,k-1}L_{liNE,cum,k} + x_{EOD,k-1}L_{liNE,cum,k-1} + x_{EOC,k}L_{liNE,cum,k} \\ &\quad - x_{EOC,k-1}L_{liNE,cum,k-1}] \\ &= N_0[(x_{EOC,k} - x_{EOD,k-1})L_{liNE,cum,k} \\ &\quad - (x_{EOC,k-1} - x_{EOD,k-1})L_{liNE,cum,k-1}]\end{aligned} \tag{S80}$$

which leads to

$$D_{slip,k} = N_0[(x_{EOC,k} - x_{EOD,k-1})L_{liNE,cum,k} - (x_{EOC,k-1} - x_{EOD,k-1})L_{liNE,cum,k-1}](1 - \lambda) \tag{S81}$$

$$Q_{loss,k} = N_0[(x_{EOC,k} - x_{EOD,k-1})L_{liNE,cum,k} - (x_{EOC,k-1} - x_{EOD,k-1})L_{liNE,cum,k-1}](1 - \lambda) \tag{S82}$$

Equations S81 and S82 are very convenient, as $L_{liNE,cum}$ rather than $L_{liNE}$ are often what is directly found experimentally. For the remaining modalities of LAM, the relevant equations in this form are:

- LAMdeNE (for when $x_{EOD,k-1} \neq x_{EOD,k-2}$)

$$C_{slip,k} = \omega N_0 \big[ (x_{EOC,k-1} - x_{EOD,k-1}) L_{deNE,cum,k} - (x_{EOC,k-1} - x_{EOD,k-2}) L_{deNE,cum,k-1} \big] \quad (S83)$$

$$Q_{loss,k} = -\omega N_0 \big[ (x_{EOC,k-1} - x_{EOD,k-1}) L_{deNE,cum,k} - (x_{EOC,k-1} - x_{EOD,k-2}) L_{deNE,cum,k-1} \big] \quad (S84)$$

- LAMliPE (for when $y_{EOD,k-1} \neq y_{EOD,k-2}$)

$$C_{slip,k} = \big[ (y_{EOC,k-1} - y_{EOD,k-2}) L_{liPE,cum,k-1} - (y_{EOC,k-1} - xy_{EOD,k-1}) L_{liPE,cum,k} \big] (1 + \omega) \quad (S85)$$

$$Q_{loss,k} = -\big[ (y_{EOC,k-1} - y_{EOD,k-2}) L_{liPE,cum,k-1} - (y_{EOC,k-1} - y_{EOD,k-1}) L_{liPE,cum,k} \big] (1 + \omega) \quad (S86)$$

- LAMdePE (for when $y_{EOC,k} \neq y_{EOC,k-1}$)

$$D_{slip,k} = \lambda \big[ (y_{EOC,k} - y_{EOD,k-1}) L_{dePE,cum,k} - (y_{EOC,k-1} - y_{EOD,k-1}) L_{dePE,cum,k-1} \big] \quad (S87)$$

$$Q_{loss,k} = \lambda \big[ (y_{EOC,k} - y_{EOD,k-1}) L_{dePE,cum,k} - (y_{EOC,k-1} - y_{EOD,k-1}) L_{dePE,cum,k-1} \big] \quad (S88)$$

Equations in Table 1 remain valid when $x$ and $y$ at the EOC and/or EOD show little variation between cycles $k$ and $k-1$. Figure S6 shows that, for the multi-modal aging assumed in Figure 6 of the main manuscript, $y_{EOC}$ varies little during the simulated cell life, indicating that LAMdePE can be reasonably described by the simplified form of the equation. However, $x_{EOC}$ presents significant variation, and most of the errors in Figure 6 and Figure 7 originate from the use of the simplified equation for LAMliNE. Figure S7 shows the same information as in Figure 6, but using the complete equations above for the calculations.

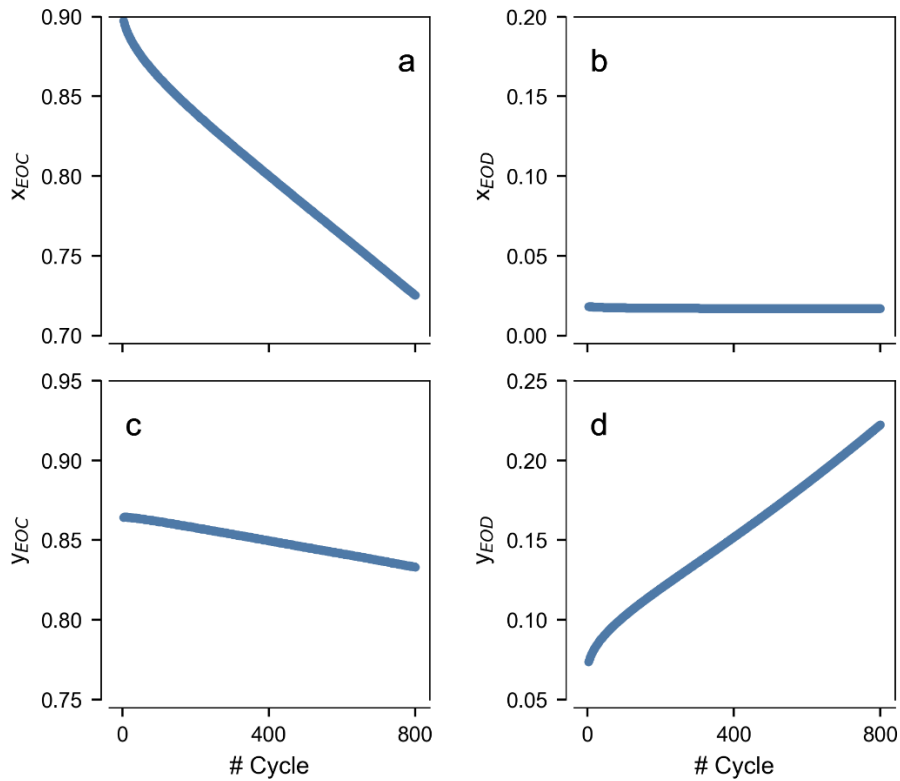

**Figure S6.** $Li^+$ content $x$ for the NE and state of delithiation $y$ for the PE at the EOC and EOD for the cell considered in Figure 6.

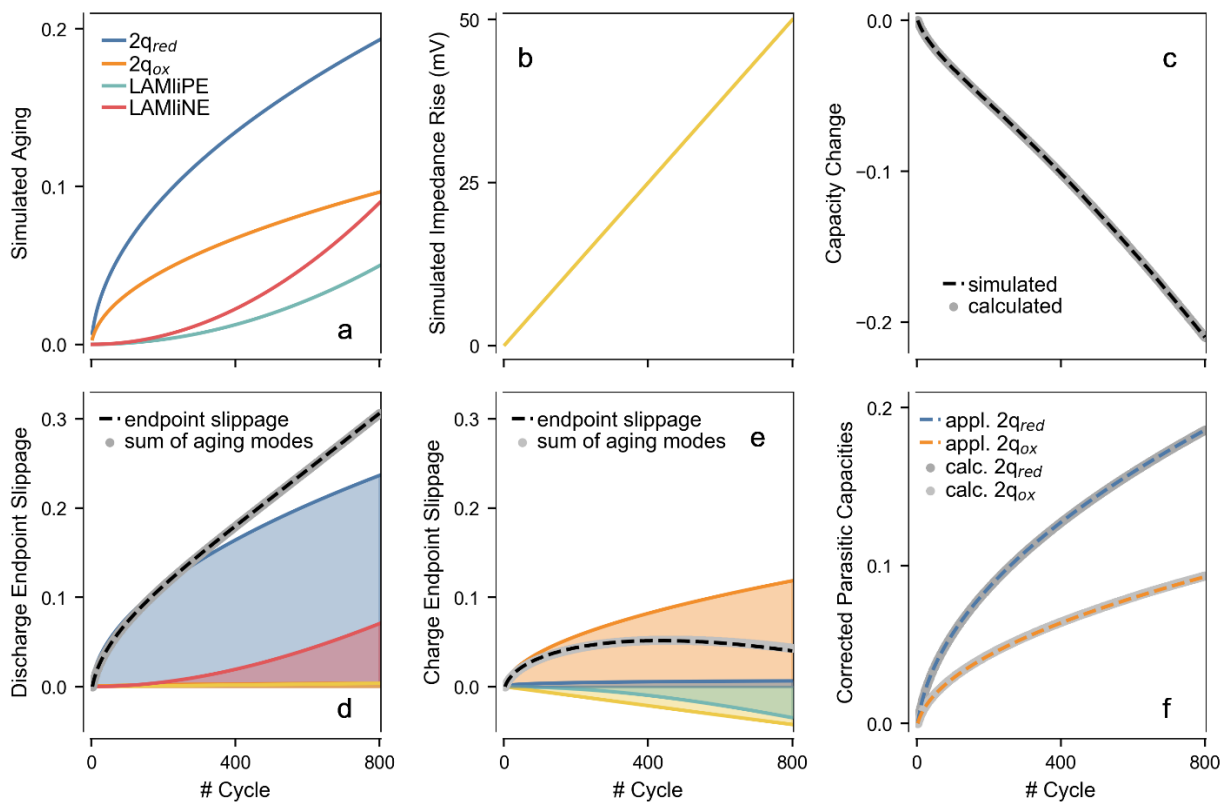

*Figure S7.* Like Figure 6 of the main manuscript, but using the equations shown in this section for calculation of endpoint slippages and capacity fade. With the more complex equations, description of aging becomes more accurate.

## *Section S8. Conversion from PE SOC to cell SOC scale*

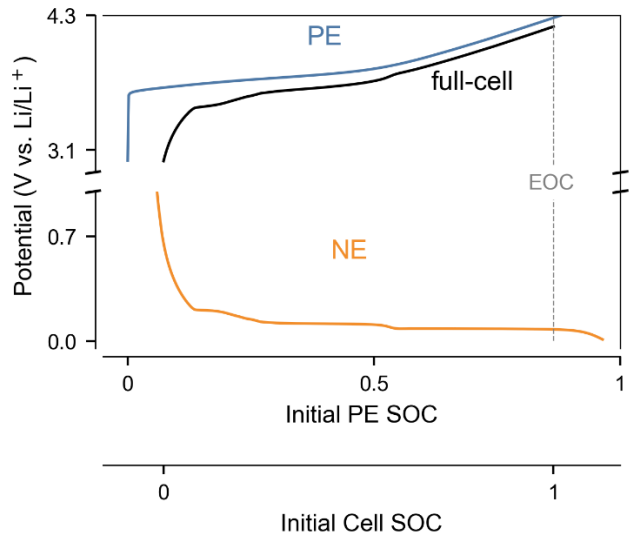

**Figure S8.** *Cycle of reference for the beginning of life of a hypothetical cell. The bottom x-axis shows the scale in the frame of reference of the full-cell. Note that portions of the PE profile at higher voltages are omitted from the figure.*

Consider an arbitrary point $y$ along the full-cell voltage profile exhibited in Figure S8. Assuming that the figure represents the initial cycle of reference, conversion of the coordinates of $y$ from the Initial PE SOC to the Initial Cell SOC scale can be achieved by simple normalization as in

$$z = \frac{y - y_{cell,min}}{y_{cell,max} - y_{cell,min}} \tag{S89}$$

where $y_{cell,min}$ and $y_{cell,max}$ are the first and last data points, respectively, for the initial cell profile in the PE SOC scale. Recalling equations S1 and S2, we see that endpoint slippages are the difference between the endpoints of cycles $k-1$ and $k$. Taking $C_{slip,k}$ as an example, we have that

$$C_{slip,k,cell} = C_{k,cell} - C_{k-1,cell} = \frac{C_k - y_{cell,min}}{y_{cell,max} - y_{cell,min}} - \frac{C_{k-1} - y_{cell,min}}{y_{cell,max} - y_{cell,min}}$$

$$= \frac{C_k - C_{k-1}}{y_{cell,max} - y_{cell,min}} = \frac{Q_{c,k} - Q_{d,k-1}}{y_{cell,max} - y_{cell,min}}$$

(S90)

The fact that endpoint slippages are the differences between two quantities eliminates the $y_{cell,min}$ term from the numerator, resulting in the conversion factor $(y_{cell,max} - y_{cell,min})^{-1}$ that can be applied to all equations in Table 1 to convert them to a cumulative cell SOC scale. For example, the endpoint slippage and capacity loss for LAMliPE becomes

$$D_{slip,k} = \frac{-L_{liPE,k}(y_{EOC,k-1} - y_{EOD,k-1})(1 + \omega)}{y_{cell,max} - y_{cell,min}}$$

(S91)

$$Q_{loss,k} = \frac{L_{liPE,k}(y_{EOC,k-1} - y_{EOD,k-1})(1 + \omega)}{y_{cell,max} - y_{cell,min}}$$

(S92)

Naturally, changing the cycle taken as reference will change the values of $y_{cell,max}$ and $y_{cell,min}$ in the conversion factor. For a well-performing cell, $y_{cell,max}$ and $y_{cell,min}$ should be approximately the same for charge and discharge, and a single conversion factor can be used for quantities related to both half-cycles.

## Section S9. Arbitrariness of x and y scales

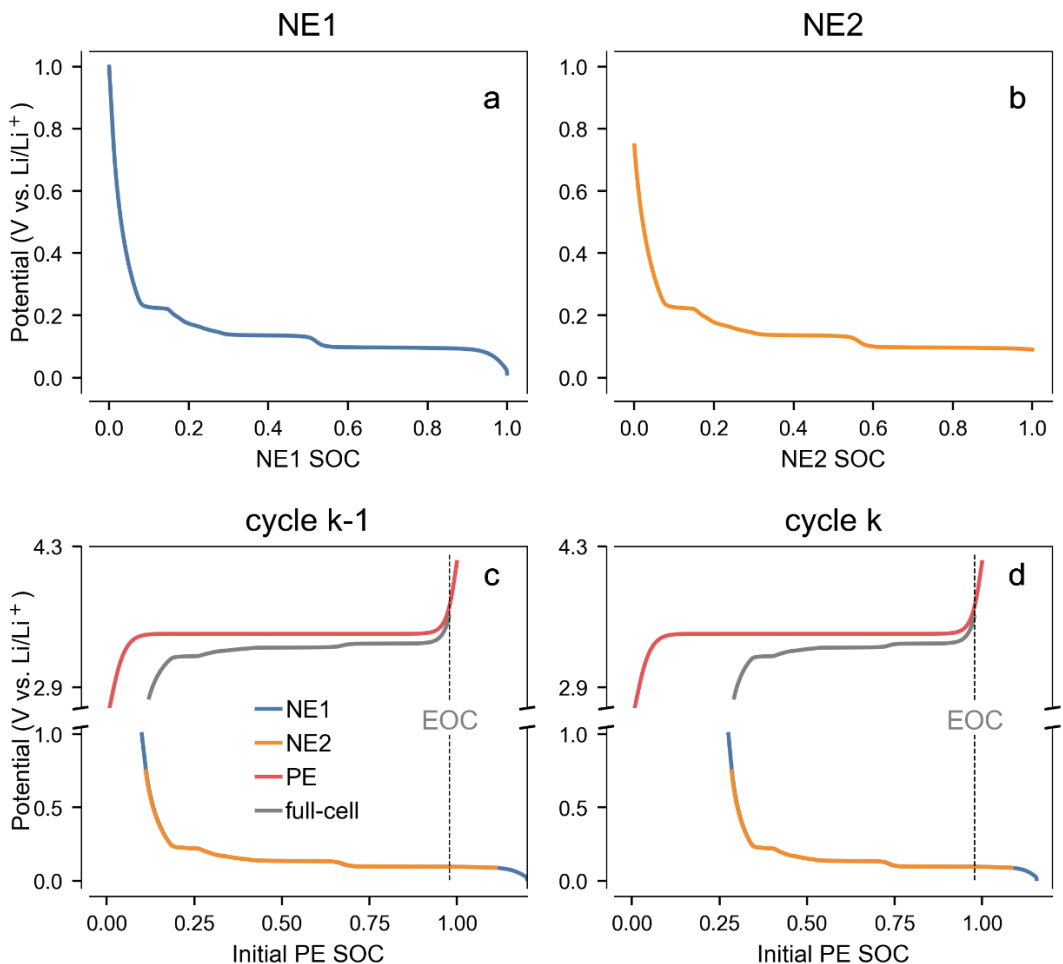

**Figure S9.** *a) Voltage profile for a graphite NE (NE1). b) Voltage profile for NE2, taken as being a section of the profile for NE1. Electrodes and full-cell charge profiles at arbitrary cycles $k-1$ (c) and $k$ (d). It is assumed that LAMliNE has occurred at the EOC of cycle $k-1$ ($LAM_{NE1} = LAM_{NE2} = 0.2$). Due to the different voltage limits of the half-cell data for each NE, a same full-cell state will correspond to different values of $x$ for each NE. For cycle $k-1$: $x_{EOC,NE1} = 0.798$, $x_{EOD,NE1} = 0.019$; $x_{EOC,NE2} = 0.867$, $x_{EOD,NE2} = 0.01$. Changing the voltage window selected for the NE will also affect the specific capacity of the electrode, requiring different N/P ratios for NE1 and NE2 to achieve the same state of the cell: $N_{0,NE1} = 1.1$; $N_{0,NE2} = 1$. With these values, we can calculate $D_{slip,k}$ caused by 20% of LAMliNE for each case to find that $D_{slip,k,NE1} = D_{slip,k,NE2} = 0.17$. That is, calculations of the endpoint slippages caused by LAM do not require absolute values of $Li^+$ content to be used.*

## Section S10. Analyzing a NMC532 vs Si cell

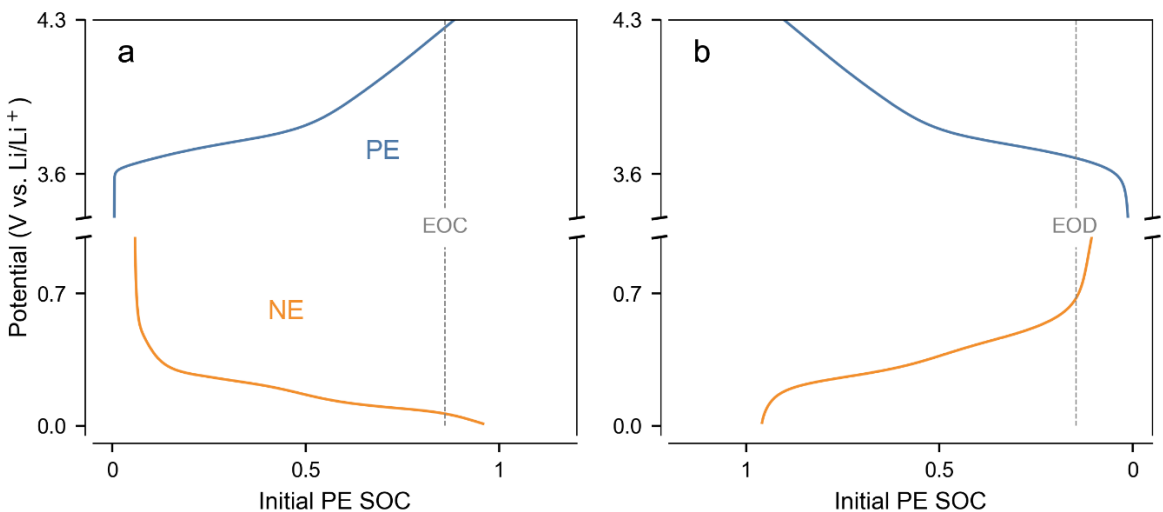

*Figure S10.* Voltage profiles for the PE and NE at the assumed initial state for the cell during: a) charge. b) discharge. The EOC and EOD points are indicated by dashed gray lines. Compare with the initial state of the graphite cell in Figure 2 of the main manuscript. Initial values of parameters of interest are $\lambda = 0.12$ and $\omega = -0.2$. Note that portions of the PE profile at higher voltages are omitted from the figure.

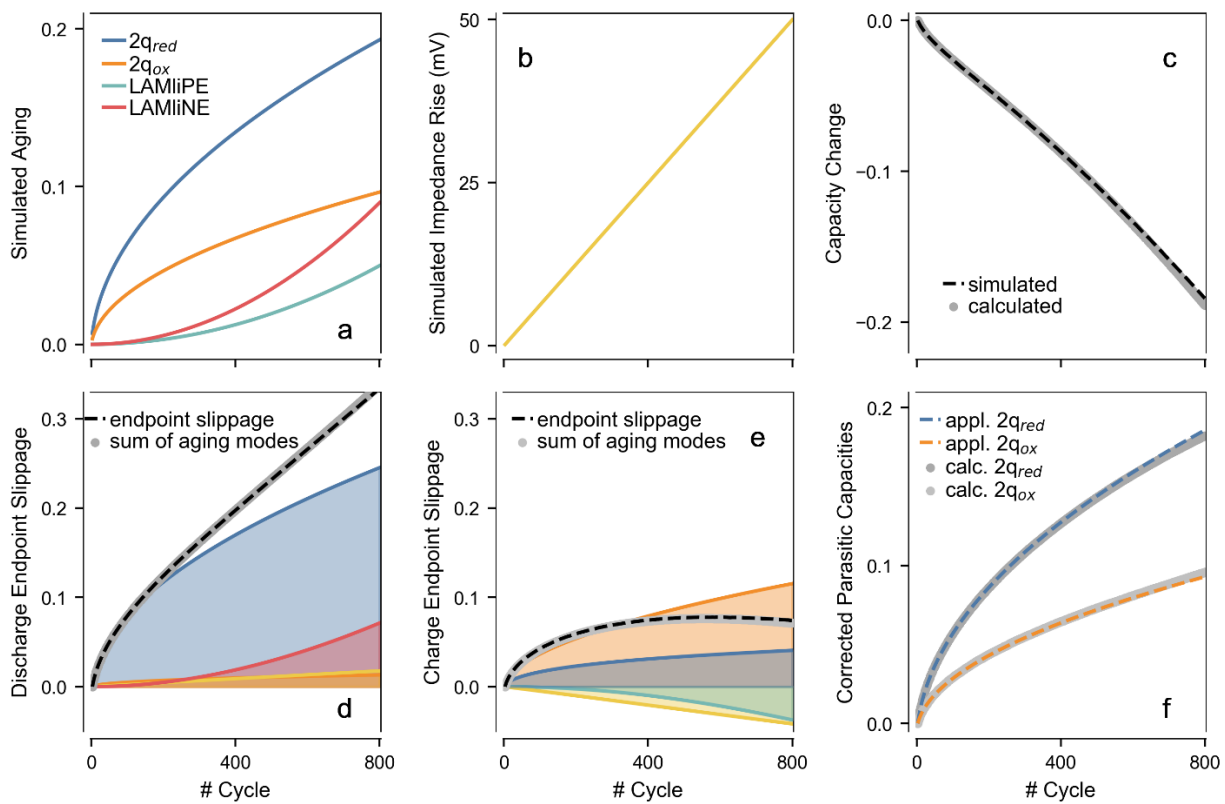

*Figure S11.* *Application of multi-modal aging to the Si cell shown in Figure S10. The aging is identical to the applied to the graphite cell in the main manuscript (Figure 6), but leads to less measurable capacity in the Si cell (panel c). Both discharge (panel d) and charge (panel e) endpoint slippages are higher in the Si cell, due to higher contributions from impedance rise and parasitic oxidation to the former, and of parasitic reduction to the latter. $q_{red}$ and $q_{ox}$ can still be accurately determined in this cell.*

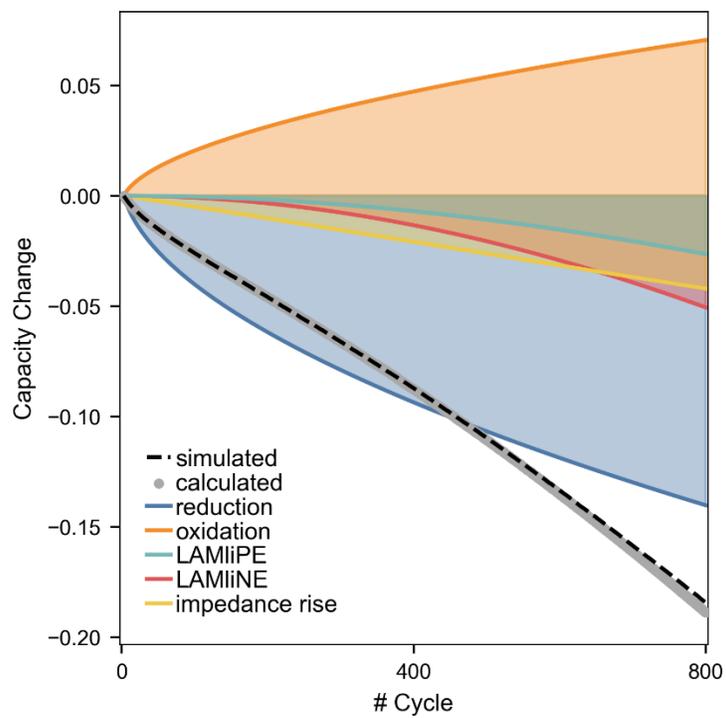

*Figure S12.* Like Figure 7, but for the Si cell shown in Figure S10. Parasitic reduction and oxidation have a lower direct contribution to capacity changes than in the case of the graphite cell, but the effects of impedance are higher.

## *Supporting References*